\documentclass[reprint,aip,jcp]{revtex4-1}
\usepackage{bbold,dcolumn,xspace,graphicx,amsmath,multirow,amssymb,xcolor,physics,verbatim,mhchem,siunitx,ragged2e}
\usepackage{simpler-wick}
\usepackage[normalem]{ulem}
\usepackage[utf8]{inputenc}

\usepackage{hyperref}
\hypersetup{
    colorlinks,
    linkcolor={red!50!black},
    citecolor={red!70!black},
    urlcolor={red!80!black}
}

\definecolor{codegreen}{rgb}{0.58,0.4,0.2}
\definecolor{codegray}{rgb}{0.5,0.5,0.5}
\definecolor{codepurple}{rgb}{0.25,0.35,0.55}
\definecolor{codeblue}{rgb}{0.30,0.60,0.8}
\definecolor{darkgreen}{rgb}{0.00,0.7,0.2}
\definecolor{backcolour}{rgb}{0.98,0.98,0.98}
\definecolor{mygray}{rgb}{0.5,0.5,0.5}

\definecolor{sqred}{rgb}{0.85,0.1,0.1}
\definecolor{sqgreen}{rgb}{0.25,0.65,0.15}
\definecolor{sqorange}{rgb}{0.90,0.50,0.15}
\definecolor{sqblue}{rgb}{0.10,0.3,0.60}

\newcommand{\hH}{\hat{H}}
\newcommand{\Htc}{\hH_{\text{TC}}}
\newcommand{\bHtc}{\boldsymbol{H}_{\text{TC}}}

\newcommand{\opad}[1]{a^{\dagger}_{#1}}
\newcommand{\duala}[1]{\tilde{a}_{#1}}
\newcommand{\dualad}[1]{\tilde{a}^{\dagger}_{#1}}

\newcommand{\rphi}[1]{{\phi_{#1}}}
\newcommand{\lphi}[1]{{\tilde{\phi}_{#1}}}
\newcommand{\rPsi}[1]{{\Psi^{(#1)}}}
\newcommand{\lPsi}[1]{{\tilde{\Psi}^{(#1)}}}
\newcommand{\rPhi}[1]{{\Phi_{#1}}}
\newcommand{\lPhi}[1]{{\tilde{\Phi}_{#1}}}

\newcommand{\br}[1]{{\boldsymbol{r}_{#1}}}

\newcommand{\bG}{\boldsymbol{G}}

\newcommand{\fnt}{\footnotetext}
\newcommand{\fnm}{\footnotemark}

\newcommand{\SupInf}{\textcolor{blue}{Supporting Information}\xspace}

\newcommand{\LCPQ}{Laboratoire de Chimie et Physique Quantiques (UMR 5626), Universit\'e de Toulouse, CNRS, UPS, France}
\newcommand{\LPT}{Laboratoire de Chimie Théorique, Sorbonne Université and CNRS, F-75005 Paris, France}

\begin{document}

\title{Compactification of Determinant Expansions via Transcorrelation}

\author{Abdallah Ammar}
\email{aammar@irsamc.ups-tlse.fr}
\affiliation{\LCPQ}
\author{Anthony Scemama}
\email{scemama@irsamc.ups-tlse.fr}
\affiliation{\LCPQ}
\author{Pierre-Fran\c{c}ois Loos}
\email{loos@irsamc.ups-tlse.fr}
\affiliation{\LCPQ}
\author{Emmanuel Giner}
\email{emmanuel.giner@lct.jussieu.fr}
\affiliation{\LPT}

\begin{abstract}
Although selected configuration interaction (SCI) algorithms can tackle much larger Hilbert spaces than the conventional full CI (FCI) method, the scaling of their computational cost with respect to the system size remains inherently exponential.
Additionally, inaccuracies in describing the correlation hole at small interelectronic distances lead to the slow convergence of the electronic energy relative to the size of the one-electron basis set.
To alleviate these effects, we show that the non-Hermitian, transcorrelated (TC) version of SCI significantly compactifies the determinant space, allowing to reach a given accuracy with a much smaller number of determinants.
Furthermore, we note a significant acceleration in the convergence of the TC-SCI energy as the basis set size increases.
The extent of this compression and the energy convergence rate are closely linked to the accuracy of the correlation factor used for the similarity transformation of the Coulombic Hamiltonian.
Our systematic investigation of small molecular systems in increasingly large basis sets illustrates the magnitude of these effects.
\end{abstract}

\maketitle

\section{Introduction}
\label{sec:introduction}
Obtaining an accurate description of the electronic structure of molecular systems remains one of the main challenges
of theoretical chemistry.
The primary challenge stems from electronic correlation effects driven by the Coulomb repulsion among electrons.
To address this issue, wave function theory appears as a promising tool as it allows for a systematically improvable solution of the Schr\"odinger equation following two complementary directions:
i) an improvement of the wave function method to get as close as possible to the full configuration interaction (FCI) solution,
and ii) an increase in the size of the one-electron basis set towards the complete basis set (CBS) limit.
These two aspects are usually considered separately and addressed by different approaches.

Regarding the basis set expansion, it was acknowledged long ago~\cite{Hyl-ZP-29} that the slow convergence
of properties with respect to the size of the one-particle basis set, computed within wave function methods,
arises from correlation effects when two electrons are close to each other, that is, near the
derivative discontinuity (the so-called electron-electron cusp)
that originates from the divergence of the Coulomb repulsion as the interelectronic distance $r_{12} \to 0$. \cite{Kat-TAMS-51,Kat-CPAM-57,PacBye-JCP-66,MyeUmrSetMor-PRA-91,MorKut-JPC-93}

To mitigate this issue, it was proposed to complement the wave function with a two-electron function (or geminal) explicitly
depending on $r_{12}$, leading to the so-called ``R12'' and the more modern ``F12'' methods. \cite{Kut-TCA-85,KutKlo-JCP-91,NogKut-JCP-94,TenNog-WIREs-12,Ten-TCA-12,HatKloKohTew-CR-12,GruHirOhnTen-JCP-17}
In the latter method, a correlation factor is included to introduce short-range correlation effects, hence improving the shape of the correlation hole at small $r_{12}$.
This leads to faster basis set convergence of ground-state properties. \cite{Sch-PR-62,KutMor-JCP-92,Hil-JCP-85,TewKlopNeiHat-PCCP-07}
Nevertheless, F12 methods do not use the full correlation factor \textit{per se} but only its orthogonal complement
with respect to the space spanned by the one-electron basis, leading to an operator that, by design,
only describes correlation effects that are absent of the one-electron basis set.
Therefore, F12 methods do not alter the expansion and potential complexity of the wave function within the basis set.

Considering now the approximations of the FCI wave function and energy,
there exists a wide variety of systematically-improvable approaches starting from the mean-field Hartree-Fock (HF) solution:
perturbation theory (PT), CI expansions, coupled-cluster (CC) theory, \cite{Ciz-JCP-66,Ciz-ACP-69,Pal-MCMP-92,Cra-RCC-00,BarMus-RMP-07,ShaBar-Book-09}
or matrix product states (MPS). \cite{Whi-PRL-92,Whi-PRB-93} Within this zoo of wave function formalisms, one can classify them into two categories:
methods for which the wave function ansatz is fixed by design (e.g., CC with single and double excitations),
and adaptive methods where the wave function automatically adapts until reaching a given accuracy.
The former has a clear advantage as their computational cost scales polynomially with the system size,
with the potential drawback that the ansatz might not always be adapted to the specific problem under study.

Among the adaptive methods, one can mention the vast family of selected configuration interaction (SCI)
\cite{BenDav-PR-69,HurMalRan-JCP-73,BuePey-TCA-74,BuePeyBru-book-81,EvaDauMal-CP-83,
Har-JCP-91,GinSceCaf-CJC-13,GinSceCaf-JCP-15,HolTubUmr-JCTC-16,SchEva-JCTC-17,GinTewGarAla-JCTC-18,
LooSceBloGarCafJac-JCTC-18,LooBogSceCafJac-JCTC-19,QP2-JCTC-19,ZhaLieuHof-JCTC-21}
approaches whose early stages originate in the late 60's, the MPS approaches \cite{ChaSha-AR-11,BaiRei-JCP-20-The} which started
in the late 90's \cite{Whi-PRL-92} or the more recent FCI quantum Monte Carlo (FCIQMC), \cite{BooThoAla-JCP-09,BooCleThoAla-JCP-11} many-body expanded FCI (MBE-FCI), \cite{EriGau-JCTC-18,EriGau-JCTC-19}
full CC reduction (FCCR), \cite{XuUejTen-PRL-18,XuUejTen-JPCL-20} and moment expansion CC [CC(P,Q)]. \cite{DeuShePie-JCP-21,GurDeuShePie-JCP-21,GurPie-JCP-23}
Although formulated in distinct ways, these adaptive methods are characterized by the idea of selecting the most
important components of the wave function among a very large Hilbert space.

One of the particularities of FCIQMC and SCI is that they both rely on a linear parametrization of the electronic wave function.
The advantage of such linear expansions is certainly the ability to rapidly and easily compute the Hamiltonian matrix
elements between the numerous Slater determinants involved in the SCI or FCIQMC calculations.
At the SCI level, this is further exploited for the computation of a second-order PT (PT2) correction,
which can be then used to extrapolate the variational energy to the FCI limit,
as originally proposed by Holmes \textit{et al.}\cite{HolUmrSha-JCP-17} and lately rationalized by Burton and Loos. \cite{BurLoo-JCP-24}
The resulting SCI+PT2 approach is now regarded as a powerful machinery to generate FCI-quality energies for both ground and excited states. \cite{ErikAndDeu-JPCL-20,LooDamSce-jCP-20,CafAplGinSce-JCP-16,HolUmrSha-JCP-17,ChiHolOttUmrShaZim-JPCA-18,LooSceBlo-JCTC-18,LooBogSceCafJac-JCTC-19,LooLipBogSceJac-JCTC-20,VerSceCafLipBogJacLoo-WIRE-21,DamVerKos-JCP-21,LooDamAmm-arXiv-24}
However, due to the intrinsic exponential nature of the exact wave function, both SCI and FCIQMC rapidly
run out of steam as the size of systems and/or basis sets increases.
Therefore, being able to compactify the CI expansion within a given basis set can substantially expand the range of applicability, in terms of system size, of these state-of-the-art approaches.

A potential solution allowing for simultaneous compactification of the CI expansion and reduction of the basis set error is to supplement the wave function by a correlation factor without projection schemes.
\cite{KutMor-JCP-92,FilUmr-JCP-96,GinAssTou-MP-16,DobLuoAla-PRB-19,Gin-JCP-21,LudDahLuc-JCTC-23}
The two main families of projection-free methods dealing with correlation factors are
the transcorrelation (TC) method of Boys and Handy, \cite{BoyHan-PRSLA-69-calculation}
which is related to earlier work by Hirschfelder, \cite{Hirschfelder-JCP-63} and the
variational Monte Carlo (VMC) approach. \cite{TouAssUmr-book-vmc}
Importantly, while VMC treats the correlation factor in a Hermitian framework, TC involves a similarity transformation
of the Hamiltonian by the same correlation factor, leading to a non-Hermitian effective Hamiltonian.

The main consequences are four-fold:
i) the VMC framework being Hermitian, it necessarily provides variational energies unlike the TC formalism;
ii) the Baker-Campbell-Hausdorff (BCH) expansion of the TC Hamiltonian
naturally truncates at second order with at most three-body terms, while the BCH expansion of
the effective VMC Hamiltonian does not truncate and therefore generates up to $N$-body terms;
iii) because TC involves at most three-body terms, the corresponding matrix elements between two Slater determinants can be computed in a deterministic way,
in contrast to VMC where one must rely on stochastic integration techniques
to compute the $N$-body integrals and avoid the curse of dimensionality;
iv) the similarity transformation of the TC Hamiltonian maintains the original overlap metric of the Slater determinant basis
while VMC necessarily introduces nonorthogonality.

Despite these differences, for a given correlation factor and in the limit of a complete basis,
the right eigenvector of the TC Hamiltonian and the Slater part of the VMC trial wave function
coincide, resulting in identical TC and VMC energies. \cite{Luo_etal_2010,Luo_2010,Luo-JCP-11,Luo_2012}
Because the TC Hamiltonian can be written in second-quantized form, it can then be employed in any post-HF methods, which, nevertheless, have to be adapted to the non-Hermitian TC formalism.
Transcorrelated versions of most wave function ans\"atze have been reported, starting from the single-determinant formalism,
\cite{BoyHan-PRSLA-69-determination,BoyHan-PRSLA-69-calculation,BoyHan-PRSLA-69-condition,
BoyHan-PRSLA-69-first,Han-JCP-69,Boy-PRSLA-69,Han-MP-71,Han-MP-72,Arm-MP-72-application,
BerBoy-MP-73,Ber-JP-73,Arm-CPL-74,SakTsu-JPCJ-75,Han-book-75,CarHyaSta-IJQC-77,HalMil-PRA-78,
Arm-JPC-80,Tsu-PTPS-08,DhaGri-PRA-13,OchSodSakTsu-JCP-12,OchSsu-JP-13,WahJimHenScu-PRB-15,OchAriTsu-PRL-17,
Ochi-CPC-23,Och-PRA-23,LeeTho-JCTC-23,AmmSceGin-JCTC-23},
to PT, \cite{Ten-CPL-00,TenHin-IJMC-02,OchTsu-CPL-15}
CI and SCI, \cite{FimUnw-IJCQ-76,OchTsu-JCTC-14,OchYamAriTsu-JCP-16,Gin-JCP-21,AmmSceGin-JCP-22,AmmGinSce-JCTC-22,AmmSceGin-JCP-23}
CC, \cite{HinTanTen-JCP-01,HinTanTen-CPL-02,Luo_2012,OchSodTsu-JCP-14,LiaSchLuoKatAla-PRR-21,SchChrRioAlaKat-JCP-23}
FCIQMC, \cite{Jeszenszki_etal_2018,LuoAla-JCTC-18,DobLuoAla-PRB-19,CohLuoGutDobTewAla-JCP-19,Jeszenszki_etal_2020,DobCohAlaGin-JCP-22}
and MPS. \cite{BaiRei-JCP-20-Transcorrelated,BaiLesRei-JCTC-22,LiaZhaChrSchRioKatAla-JCTC-23}

As mentioned in previous studies,
\cite{KutMor-JCP-92,FilUmr-JCP-96,GinAssTou-MP-16,DobLuoAla-PRB-19,Gin-JCP-21,LudDahLuc-JCTC-23}
the inclusion of a correlation factor, either within the TC or VMC framework,
allows to compactify the determinantal part of the wave function.
This key feature could potentially expand the applicability of SCI while decreasing the finite basis set error.
Moreover, the superior quality of TC energy differences has been previously reported in several studies.
\cite{CohLuoGutDobTewAla-JCP-19,LiaSchLuoKatAla-PRR-21,AmmSceGin-JCP-22,AmmSceGin-JCP-23,SchChrRioAlaKat-JCP-23,LiaZhaChrSchRioKatAla-JCTC-23}
The aim of the present work is to focus on the improved convergence properties of TC-SCI with respect to the standard SCI implementation for both absolute and relative energies.
More precisely, we aim to study the impact of the correlation factor on three key aspects of the SCI algorithm:
i) the convergence of the TC-SCI energy, ii) the selection of the Slater determinants, and iii) the choice of the orbital set.

The paper is organized as follows. In Sec.~\ref{sec:theory}, we recall the theoretical background
of the various methods employed here. The general theory of TC is summarized in Sec.~\ref{sec:tc_theory}.
Then, the biorthogonal basis set representation of the TC Hamiltonian is presented in Sec.~\ref{sec:bio_theory}
and its application to the orbital optimization in Sec.~\ref{sec:tc_scf_theory}.
The working equations for the normal-ordering approximation of the three-body interaction are gathered in
Sec.~\ref{sec:nol_theory}, the general TC-SCI algorithm is presented in Sec.~\ref{sec:sci_theory},
and the correlation factor is described in Sec.~\ref{sec:jastrow_theory}.
In Sec.~\ref{sec:integral_theory}, we discuss the computation of the various additional integrals required within
the TC formalism, while Sec.~\ref{sec:comp_cost} analyses the main computational bottleneck associated with TC calculations.
Then, Sec.~\ref{sec:results} presents the main numerical results for atomic and molecular systems.
Sec.~\ref{sec:comp_det} gathers the computational details.
Considering the water molecule as an example, we carry out a detailed investigation of
i) the role of the correlation factor and orbital optimization on the distribution of the determinant weights in the
TC-SCI expansion (Sec.~\ref{sec:conv_weight}), ii) the convergence of the non-variational TC-SCI energy and its extrapolation towards the FCI limit (Sec.~\ref{sec:conv_extrap}), and
iii) the compactification of the determinant expansion provided by the TC formalism (Sec.~\ref{sec:compact}).
We also study, in Sec.~\ref{sec:total_energies} and Sec.~\ref{sec:IP}, the convergence of total energies
and ionization potentials (IPs) obtained within the TC-SCI approach.
Comparison with the best estimates from the literature and CBS values are reported.
Finally, our conclusions are drawn in Sec.~\ref{sec:conclusion}.
Unless otherwise stated, atomic units are used throughout.

\section{Theoretical Framework}
\label{sec:theory}
From hereon, the indices $p,q,r,s,\ldots$ denote arbitrary spin-orbitals, while
$i,j,k,\ldots$ and $a,b,c,\ldots$ designate occupied and virtual spin-orbitals, respectively.
Moreover, the indices $\mu,\nu,\lambda,\sigma,\ldots$ represent basis functions, i.e., atomic orbitals.

\subsection{Transcorrelated Approach}
\label{sec:tc_theory}
The central concept of the TC theory is to apply a similarity transformation
\begin{equation}
	\Htc \equiv e^{-\hat{\tau}} \hH e^{\hat{\tau}}
	\label{eq:Htc_def}
\end{equation}
to the bare Coulombic Hamiltonian
\begin{equation}
	\hH = - \sum_i \frac{\nabla_i^2}{2} - \sum_i \sum_A  \frac{Z_A}{r_{iA}} + \sum_{i<j} \frac{1}{r_{ij}},
	\label{eq:H_def}
\end{equation}
where $\br{i}$ represents the position of electron $i$, the index $A$ runs over the nuclei of charge $Z_A$, $r_{iA}$ is the distance between the $i$th electron and the $A$th nucleus, and $r_{ij} = \abs{\br{i} - \br{j}}$.
The similarity transformation defined in Eq.~\eqref{eq:Htc_def} involves a correlation (or Jastrow) factor
\begin{equation}
	\hat{\tau} \equiv \sum_{i<j} u_{ij},
\end{equation}
where the two-electron function $u_{ij} \equiv u \qty(\br{i},\br{j})$ is symmetric with respect to the exchange of two
electrons, i.e., $u_{ji} = u_{ij}$.

The two Hamiltonians, $\hH$ and $\Htc$, defined in Eqs.~\eqref{eq:H_def} and \eqref{eq:Htc_def} respectively, share the same spectrum of eigenvalues
in a complete basis.
The similarity transformation allows the transfer of physical effects from the correlation factor to the TC Hamiltonian.
Therefore, by incorporating correlation effects and exact conditions in $\hat{\tau}$, such as the electron-electron cusp,
one would anticipate the eigenvalues and eigenvectors of $\Htc$ to converge faster toward the CBS limit when employing finite basis sets.

Using the BCH expansion, which here truncates naturally at second order, one can show that the TC Hamiltonian reads
\begin{equation}
	\Htc
	=
	\hH  -
        \sum_{i<j} \hat{K}_{ij} \\
	-
        \sum_{i<j<k} \hat{L}_{ijk},
\end{equation}
where the two-electron operator $\hat{K}_{ij}$ and the three-electron operator $\hat{L}_{ijk}$ are explicitly given by
\begin{equation}
\begin{split}
	\hat{K}_{12}
	&\equiv
	\qty(\grad_1 u_{12}) \cdot \grad_1 + \qty(\grad_2 u_{21}) \cdot \grad_2 \\
	&+
	\frac{1}{2} \grad_1^2 u_{12} + \frac{1}{2} \grad_2^2 u_{21} \\
	&+
	\frac{1}{2} \qty(\grad_1 u_{12})^2 + \frac{1}{2} \qty(\grad_2 u_{21})^2,
\end{split}
	\label{eq:Kop_def}
\end{equation}
and
\begin{equation}
\begin{split}
	\hat{L}_{123}
	&\equiv
	\qty(\grad_1 u_{12}) \cdot \qty(\grad_1 u_{13})
	\\
	&+
	\qty(\grad_2 u_{21}) \cdot \qty(\grad_2 u_{23})
	\\
	&+
	\qty(\grad_3 u_{32}) \cdot \qty(\grad_3 u_{31}).
\end{split}
	\label{eq:Lop_def}
\end{equation}
Clearly, $\Htc$ is non-Hermitian and includes three-electron operators. These notable
features will be thoroughly examined and addressed in the subsequent sections.

\subsection{Biorthogonal Representation}
\label{sec:bio_theory}

Given that the eigenvectors of a non-Hermitian Hamiltonian are generally not orthogonal, the biorthogonal
framework \cite{Moshinsky_Seligman_1971,Brody_2014} emerges as the most natural approach to handle $\Htc$. \cite{FimUnw-IJCQ-76,TenHin-IJMC-02,OchTsu-CPL-15,AmmSceGin-JCP-23,AmmSceGin-JCTC-23,Ochi-CPC-23,Och-PRA-23,LeeTho-JCTC-23}
By introducing two sets of one-electron functions, $\{\lphi{p}\}$ and $\{\rphi{p}\}$, satisfying the biorthogonality condition
\begin{equation}
	\braket*{\lphi{p}}{\rphi{q}} = \delta_{pq},
	\label{eq:biorthog_cond}
\end{equation}
one can define direct and dual creation and annihilation operators by their actions on the true vacuum $\ket{0}$, as follows: \cite{Moshinsky_Seligman_1971}
\begin{align}
	\opad{p} \ket{0} &= \ket*{\rphi{p}}, &
	\dualad{p} \ket{0} &= \ket*{\lphi{p}}.
\end{align}
The biorthogonal condition~\eqref{eq:biorthog_cond} ensures that the creation and annihilation operators
satisfy the anticommutation relations within the context of biorthogonality, that is,
\begin{align}
	\{ \duala{p},  \opad{q} \} &= \delta_{pq}, &
	\{  \opad{p},  \opad{q} \} &= 0, &
	\{ \duala{p}, \duala{q} \} &= 0.
        \label{eq:anticomm_dual}
\end{align}
Hence, in the second-quantization formalism, $\Htc$ can be decomposed as a sum of one-, two-, and three-body terms, with explicit forms
\begin{equation}
\begin{split}
	\Htc
	&=
	\sum_{pq} h_{q}^{p} \opad{p} \duala{q} \\
	&+
	\frac{1}{2!} \sum_{pqrs} V_{rs}^{pq} \opad{p} \opad{q} \duala{s} \duala{r} \\
	&+
	\frac{1}{3!} \sum_{pqrstu}
		L_{stu}^{pqr} \, \opad{p} \opad{q} \opad{r} \duala{u} \duala{t} \duala{s},
\end{split}
\end{equation}
where
\begin{subequations}
\begin{align}
	\label{eq:1e_integ}
	h_{q}^{p} &= \mel*{\lphi{p}}{\hat{h}}{\rphi{q}}, \\
	\label{eq:2e_integ}
	V_{rs}^{pq} &= \mel*{\lphi{p} \lphi{q}}{r_{12}^{-1} - \hat{K}_{12}}{\rphi{r} \rphi{s}}, \\
	\label{eq:3e_integ}
	L_{stu}^{pqr} &= \mel*{\lphi{p} \lphi{q} \lphi{r}}{-\hat{L}_{123}}{\rphi{s} \rphi{t} \rphi{u}}.
\end{align}
\end{subequations}
%

\subsection{Self-Consistent Field Procedure}
\label{sec:tc_scf_theory}
Unless one aims to perform a FCI calculation, the choice of the underlying orbitals considered to build the multideterminant expansion for a given approximated correlated treatment may have a significant impact on the overall accuracy and convergence properties.
In the case of standard ground-state SCI calculations, one usually relies on mean-field HF orbitals as a starting point which guarantees the fulfilment of Brillouin's theorem.
Regarding now TC-SCI calculations, as $\Htc$ differs from the bare Coulomb Hamiltonian, the usual HF orbitals are
not optimal with respect to the TC effective potential.
Relying on non-optimal orbitals may impact the computation of energy differences when enforcing the versatile frozen-core approximations, \cite{AmmSceGin-JCP-23}
and can also deteriorate the convergence of TC-SCI calculations as we shall discuss in Sec.~\ref{sec:conv_extrap}.

To address these challenges, the Brillouin theorem is generalized to the non-Hermitian case,
\cite{FimUnw-IJCQ-76,TenHin-IJMC-02,OchTsu-CPL-15,AmmSceGin-JCTC-23,Ochi-CPC-23,Och-PRA-23,LeeTho-JCTC-23}
which implies two distinct left and right conditions.
In the case of a closed-shell system, Brillouin's theorem translates into finding
a pair of Slater determinants, $\lPhi{0}$ and $\rPhi{0}$, built with the occupied orbitals of the sets
$\{\lphi{p}\}$ and $\{\rphi{p}\}$, respectively, fulfilling the following conditions:
\begin{subequations}
\begin{align}
  \label{eq:brillouin_left}
  \mel*{\lPhi{0}}{\opad{i}\duala{a} \Htc}{\rPhi{0}} &= 0,
  \\
  \label{eq:brillouin_right}
  \mel*{\lPhi{0}}{\Htc \opad{a}\duala{i}}{\rPhi{0}} &= 0.
\end{align}
\end{subequations}
The sets of orbitals $\{\lphi{p}\}$ and $\{\rphi{p}\}$ are necessarily biorthogonal and can be obtained through
the TC self-consistent field (SCF) procedure as detailed in Ref.~\onlinecite{AmmSceGin-JCTC-23}.

This non-linear optimization process is carried out iteratively by
diagonalizing the non-symmetric TC Fock matrix, which reads
\begin{equation}
\label{eq:tc_fock}
\begin{aligned}
	F_q^p
	&\equiv
	h_q^p + \sum_{i} \bar{V}_{i q}^{i p} + \frac{1}{2} \sum_{ij} \bar{L}_{i j q}^{i j p},
\end{aligned}
\end{equation}
where we have introduced the antisymmetrized set of integrals
\begin{subequations}
\begin{align}
	\bar{V}_{rs}^{pq} &\equiv V_{r s}^{p q} - V_{s r}^{p q}, \\
	\bar{L}_{s t u}^{p q r} & \equiv
		L_{s t u}^{p q r} - L_{s u t}^{p q r} + L_{t u s}^{p q r}
		- L_{t s u}^{p q r} + L_{u s t}^{p q r} - L_{u t s}^{p q r}.
\end{align}
\end{subequations}
At convergence of the TC-SCF procedure, the left and right TC-SCF orbitals are the left and right eigenvectors of the non-Hermitian TC Fock matrix defined in Eq.~\eqref{eq:tc_fock}.

\subsection{Normal-Ordering with Biorthogonal Orbitals}
\label{sec:nol_theory}
Storing three-electron integrals poses a significant computational burden, demanding extensive memory resources in practice.
To handle the three-electron interaction term in $\Htc$ (see Sec.~\ref{sec:tc_theory}),
we apply the normal-ordering technique to the three-body operator
\begin{equation}
  \hat{L}=\frac{1}{3!} \sum_{pqrstu}
		L_{stu}^{pqr} \opad{p} \opad{q} \opad{r} \duala{u} \duala{t} \duala{s},
\end{equation}
written in a biorthogonal basis,
thus generalizing recent work performed in the context of an orthogonal basis. \cite{ChrSchLopAlaKat-JCP-23}

In our case, the Slater determinants $\lPhi{0}$ and $\rPhi{0}$ obtained through the previous TC-SCF procedure are defined as the Fermi vacua.
More specifically, we have
\begin{equation}
\begin{aligned}
	\opad{p} \duala{q}
	&=
	\mathcal{N}\qty[\opad{p} \duala{q}] + \wick{\c{\opad{p}} \c{\duala{q}}}
\end{aligned}
\end{equation}
where
\begin{equation}
\label{eq:gamma}
\begin{aligned}
	\wick{\c{\opad{p}} \c{\duala{q}}}
	\equiv \mel*{\lPhi{0}}{\opad{p} \duala{q}}{\rPhi{0}}
	= \gamma_{q}^{p},
\end{aligned}
\end{equation}
and the normal-ordered term is defined such that
\begin{equation}
	\mathcal{N}[\opad{p} \opad{q} \cdots \duala{r} \duala{s} \cdots] \equiv 0.
\end{equation}
In Eq.~\eqref{eq:gamma}, one can recognize the transition density matrix $\gamma_{q}^{p}$ which is the equivalent of the
usual density matrix in Hermitian calculations.

By applying the following identity
\begin{equation}
\begin{aligned}
	\opad{p} \opad{q} \opad{r} \duala{u} \duala{t} \duala{s}
	&=
	\mathcal{N}\qty[\opad{p} \opad{q} \opad{r} \duala{u} \duala{t} \duala{s}]\\
	&
	+\mathcal{N}\qty[\wick{\c{\opad{p}} \opad{q} \opad{r} \c{\duala{u}} \duala{t} \duala{s}}]
	+ \cdots \\
	&
	+\mathcal{N}\qty[\wick{\c1{\opad{p}} \c2{\opad{q}} \opad{r} \c1{\duala{u}} \c2{\duala{t}} \duala{s}}]
	+ \cdots \\
	&
	+\mathcal{N}\qty[\wick{\c1{\opad{p}} \c2{\opad{q}} \c3{\opad{r}} \c1{\duala{u}} \c2{\duala{t}} \c3{\duala{s}}}]
	+ \cdots,
\end{aligned}
\end{equation}
considering all possible single, double, and triple contractions, along with the following relationships
\begin{subequations}
\begin{align}
	\mathcal{N}\qty[\opad{p} \duala{q}]
	& = \opad{p} \duala{q} - \gamma_{q}^{p}, \\
\begin{split}
	\mathcal{N}\qty[\opad{p} \opad{q} \duala{s} \duala{r}]
	&=
	\opad{p} \opad{q} \duala{s} \duala{r} \\
	&
	-\gamma_{r}^{p} \opad{q} \duala{s} + \gamma_{r}^{q} \opad{p} \duala{s}
	-\gamma_{s}^{q} \opad{p} \duala{r} + \gamma_{s}^{p} \opad{q} \duala{r} \\
	&
	-\gamma_{r}^{q} \gamma_{s}^{p} + \gamma_{s}^{q} \gamma_{r}^{p}.
\end{split}
\end{align}
\end{subequations}
one ends up with the following form for the three-body operator:
\begin{equation}
	{\hat{L}}
	= \hat{L}^{(0)} + \hat{L}^{(1)} + \hat{L}^{(2)} + \hat{L}^{(3)},
\end{equation}
with
\begin{subequations}
\begin{align}
	\label{eq:L0}
	\hat{L}^{(0)} &= \frac{1}{6} \sum_{i j k}
		\qty( L_{i j k}^{i j k} + 2 L_{k i j}^{i j k} - 3 L_{k j i}^{i j k} ), \\
	\label{eq:L1}
	\hat{L}^{(1)} &= \sum_{pq} \tilde{L}_{q}^{p}  \opad{p} \duala{q}, \\
	\label{eq:L2}
	\hat{L}^{(2)} &= \frac{1}{2} \sum_{pqrs} \tilde{L}_{rs}^{pq} \opad{p} \opad{q} \duala{s} \duala{r}, \\
	\hat{L}^{(3)} &= \frac{1}{6} \sum_{pqrstu} L_{stu}^{pqr}
		\mathcal{N} \qty[\opad{p} \opad{q} \opad{r} \duala{u} \duala{t} \duala{s}],
\end{align}
\end{subequations}
and
\begin{subequations}
\begin{align}
	\label{eq:l1}
        \tilde{L}_{q}^{p} &= -\frac{1}{2}\sum_{i j} \qty( L_{q i j}^{p i j} - L_{q j i}^{p i j} + 2 L_{i j q}^{p i j} - 2 L_{j i q}^{p i j} )  ,\\
	\label{eq:l2}
        \tilde{L}_{rs}^{pq} &=\sum_i \qty( L_{i r s}^{i p q} - L_{r i s}^{i p q} - L_{s r i}^{i p q} ),
\end{align}
\end{subequations}
where we have used the fact that, in the case of single-determinant Fermi vacua, the only non-zero elements of the transition density matrix are $\gamma_i^i = 1$.
Therefore, the three-body operator can be written as a sum of a scalar quantity, $\hat{L}^{(0)}=\mel*{\lPhi{0}}{\hat{L}}{\rPhi{0}}$, one- and two-body operators, $\hat{L}^{(1)}$ and $\hat{L}^{(2)}$, together with a
three-body operator involving normal-ordered annihilation and creation operators.
Since, by definition, $\mel*{\lPhi{0}}{\hat{L}^{(3)}}{\rPhi{0}} = 0$, a reasonable assumption is that
the effect of the pure three-body normal-ordered operator $\hat{L}^{(3)}$ is negligible. Therefore, the normal-ordered
approximation,
\begin{equation}
 \label{eq:normal_approx}
	\hat{L} \approx \hat{L}^{(0)} + \hat{L}^{(1)} + \hat{L}^{(2)},
\end{equation}
typically results in small errors and effectively reduces the TC Hamiltonian to a two-body operator
\begin{equation}
	\Htc
	=
	\hat{L}^{(0)} +
	\sum_{pq} \tilde{h}_{q}^{p} \opad{p} \duala{q} +
	\frac{1}{2!} \sum_{pqrs} \tilde{V}_{rs}^{pq} \opad{p} \opad{q} \duala{s} \duala{r},
\end{equation}
with new effective one- and two-body integrals defined as
\begin{align}
\label{eq:h_tc_nol}
 \tilde{h}_{q}^{p} & = {h}_{q}^{p} + \tilde{L}_{q}^{p}, &
 \tilde{V}_{rs}^{pq} & = V_{rs}^{pq} + \tilde{L}_{rs}^{pq},
\end{align}
where $\tilde{L}_{q}^{p}$ and $\tilde{L}_{rs}^{pq}$ are given by Eqs.~\eqref{eq:l1} and \eqref{eq:l2}, respectively.

\subsection{Selected Configuration Interaction Algorithm}
\label{sec:sci_theory}
In the TC framework, the optimal ground-state coefficients for a selected space of Slater determinants, referred to as the SCI wave function,
are simply the ground-state left and right eigenvectors of the matrix representation of $\Htc$ within this space.
This approach offers several advantages over other optimization schemes, such as those based on the expensive and noisy VMC method,
making the TC theory highly appealing.
Furthermore, an intriguing possibility is to go a step further and directly tailor the space of determinants to adapt
to the correlation factor. This can be achieved through TC-SCI algorithms, which extend standard SCI versions.

The \textit{``Configuration Interaction using a Perturbative Selection made Iteratively''} (CIPSI) algorithm~\cite{HurMalRan-JCP-73} is highly efficient for constructing the space of determinants and providing accurate
estimates of the FCI energy with compact wave functions.
Recently, we extended the CIPSI method to accommodate the TC Hamiltonian. \cite{AmmSceGin-JCP-22,AmmSceGin-JCP-23}
In this section, we outline the TC-CIPSI algorithm, summarizing its key steps as follows:
\begin{enumerate}

\item Begin with a zeroth-order wave function
$\ket*{\rPsi{0}} = \sum_{I \in \mathcal{I}} c_I \ket*{D_I}$ within a selected space of
determinants $\mathcal{I}$, along with its dual
$\bra*{\lPsi{0}} = \sum_{I \in \mathcal{I}} \tilde{c}_I \bra*{\tilde{D}_I}$, satisfying the following conditions:
\begin{subequations}
\label{eq:E0TC}
\begin{align}
        \bHtc\ket*{\rPsi{0}} & = E_{\text{TC}}^{(0)}\ket*{\rPsi{0}}, \\
        \bHtc^\dagger \ket*{\lPsi{0}} &= E_{\text{TC}}^{(0)}\ket*{\lPsi{0}}, \\
	\braket*{\lPsi{0}}{\rPsi{0}} &= 1,
\end{align}
\end{subequations}
where $\bHtc$ is the matrix representation of $\Htc$ in the
biorthogonal basis of selected determinants.

\item Compute the energetic contributions of all determinants outside $\mathcal{I}$
\begin{equation}
 \label{eq:second_e}
	e_\alpha^{(2)} = \frac{\mel*{\lPsi{0}}{\Htc}{D_\alpha} \mel*{\tilde{D}_\alpha}{\Htc}{\rPsi{0}}}{{E_{\text{TC}}^{(0)}} - \mel*{\tilde{D}_\alpha}{\Htc}{D_\alpha}},
\end{equation}
which all together form the PT2 energy
\begin{equation}
 \label{eq:second_e_tot}
	E_{\text{TC}}^{(2)} = \sum_{\alpha \notin \mathcal{I}} e_\alpha^{(2)}.
\end{equation}
This second-order perturbative energy is computed using an efficient stochastic implementation, as proposed
in Ref.~\onlinecite{GarSceLooCaf-JCP-17}.
Because we rely on the normal-ordering approximation of Eq.~\eqref{eq:normal_approx}, $\Htc$ remains a two-body operator,
and therefore, in practice, the determinants $\ket*{D_\alpha}$ and $\ket*{\tilde{D}_\alpha}$ for which $e_\alpha^{(2)} \neq 0$
correspond to the singly- and doubly-excited determinants with respect to any selected determinant, as in
usual SCI calculations.

\item Choose a new ensemble of determinants $\mathcal{A}$, identified by the largest
energy contributions $\abs*{e_\alpha^{(2)}}$. Furthermore, when opting for a particular determinant, all other determinants
within the corresponding configuration state function are automatically incorporated,
ensuring pure spin states. \cite{ChiAppGasLooSce-AQC-21}

\item Update the zeroth-order space $\mathcal{I} \leftarrow \mathcal{I} \cup \mathcal{A}$, and apply a nonsymmetric
Davidson algorithm \cite{HirNak-JCP-82} to update $\lPsi{0}$, $\rPsi{0}$, and $E_{\text{TC}}^{(0)}$.

\item Repeat steps 2-4 until convergence is achieved.

\end{enumerate}

We would like to highlight some of the subtleties of the TC-SCI algorithm, that arise from the non-Hermitian character of $\Htc$.
First, regarding the set of determinants $\{D_I\}$ and $\{\tilde{D}_I\}$,
it is important to note that both are constructed from determinants with identical \textit{orbital occupancies}.
As a consequence, when working with a unique set of orthogonal orbitals, $\{D_I\}$ and $\{\tilde{D}_I\}$ are strictly identical,
whereas, when using a couple of biorthogonal orbital sets, they differ as each set of determinants is built
with a different set of orbitals.
Nevertheless, the biorthogonality relation between orbitals implies a biorthogonality relation between
the determinants
(see Ref.~\onlinecite{AmmSceGin-JCP-23} for a more detailed discussion).
Second, as shown in Ref.~\onlinecite{AmmSceGin-JCP-22}, the use of $e_\alpha^{(2)}$ [see Eq.~\eqref{eq:second_e}] corresponds
to a balanced selection criterion for both the left  and right eigenvectors as it involves both the left  and right zeroth-order
wave functions $\lPsi{0}$ and $\rPsi{0}$.
Third, the use of the absolute value of $e_\alpha^{(2)}$ as a selection criterion is mandatory as the expression
of Eq.~\eqref{eq:second_e} is not necessarily negative, contrary to the Hermitian case.

To conclude this section, we briefly present the extrapolation scheme used in the context of TC-SCI,
allowing us to estimate the TC-FCI energy. \cite{AmmSceGin-JCP-23}
The procedure is the direct application of the scheme initially proposed in Ref.~\onlinecite{HolUmrSha-JCP-17} for
the Hermitian case.
For small enough $E_{\text{TC}}^{(2)}$ values, one can approximate the TC-FCI energy by
\begin{equation}
 \label{eq:pt2}
 E_{\text{TC-FCI}} \approx E_{\text{TC}}^{(0)} + E_{\text{TC}}^{(2)},
\end{equation}
the TC-FCI limit being effectively and rigorously reached when $E_{\text{TC}}^{(2)} = 0$.
This implies that, as $E_{\text{TC}}^{(2)} \to 0$, the zeroth-order energy $E_{\text{TC}}^{(0)}$ becomes a linear
function of the second-order
energetic correction $E_{\text{TC}}^{(2)}$.
At each TC-SCI iteration, we compute both $E_{\text{TC}}^{(0)}$ and $E_{\text{TC}}^{(2)}$, and when the
linear regime is reached $E_{\text{TC-FCI}}$ is estimated by linearly fitting the former as a function of the latter.

The main difference with the usual SCI extrapolation is the presence of
both positive and negative contributions in $E_{\text{TC}}^{(2)}$, which implies that
the latter is not necessarily negative. Therefore,
$E_{\text{TC}}^{(0)}$ is not necessarily a monotonic function of $E_{\text{TC}}^{(2)}$.
In some cases, these two aspects can lead to situations where it is difficult to estimate faithfully the
TC-FCI energy, as shall be exemplified below.



\subsection{Correlation Factor}
\label{sec:jastrow_theory}

Various forms of correlation factors have been proposed in the literature, from the simplest universal two-body
correlation factor based on Gaussian geminals~\cite{Boy-PRSLA-60,LonSin-TCA-64,TenNo-CPL-00-a} to more elaborate forms
including electron-electron-nucleus terms.~\cite{BoyHan-PRSLA-69-calculation,UmrWilWil-PRL-88,FilUmr-JCP-96,AmmSceGin-JCTC-23}
In the present investigation, we consider the correlation factor proposed by Boys and Handy, \cite{BoyHan-PRSLA-69-calculation} which we re-express as follows:
\begin{equation}
\label{eq:general_j}
        u_{ij}
        =
        \sum_{A} \sum_{p_A=1}^{P_A} C_{m_{p_A}}^{ n_{p_A}} c_{p_A} \qty[f_{\alpha_A}(r_{ij})]^{\ell_{p_A}}
        g_{\beta_A}^{m_{p_A}n_{p_A}}(r_{iA},r_{jA})
\end{equation}
with
\begin{subequations}
\begin{align}
\label{eq:f_h_j}
        f_{\alpha}(x)       &=\frac{\alpha x}{1 + \alpha x},  \\
        g_{\alpha}^{mn}(x,y) & = \qty[f_{\alpha}(x)]^m \qty[f_{\alpha}(y)]^n + \qty[f_{\alpha}(y)]^m \qty[f_{\alpha}(x)]^n .
\end{align}
\end{subequations}
For each nucleus, we employ a total of $P_A$ terms, each term $p_A$ being
characterized by three positive integers $(\ell_{p_A},m_{p_A},n_{p_A})$ and a coefficient $c_{p_A}$.

%
Following the work of Schmidt and Moskowitz, \cite{SchMos-JCP-90} we set $\alpha_A = \beta_A =1$
for all nuclei. Additionally, we have
\begin{equation}
        C_m^n =
        \begin{cases}
                1/2,  & \text{if $m=n$},
                \\
                1,    & \text{otherwise}.
        \end{cases}
\end{equation}
Furthermore, we enforce the electron-electron cusp by systematically including in the parametrization,
the term characterized by $m_1=n_1=0$ and $\ell_1=1$, with the corresponding coefficient $c_{1}=1/2$
(see Ref.~\onlinecite{SchMos-JCP-90} for more details).

It is important to highlight the three complementary effects introduced by the parametrization of Eq.~\eqref{eq:general_j}.
First, the terms characterized by $\ell > 0$ and $m=n=0$ correspond to pure electron-electron terms,
which explicitly depend on the interelectronic distances.
Their main purpose is to model the correlation hole, that is, the lowering of the probability of finding two electrons close together due to electron correlation effects.
Second, for $\ell = m = 0$ and $n > 0$ (or $\ell = n = 0$ and $m > 0$), one introduces pure electron-nucleus terms
which provide flexibility to the one-electron density for its adjustment to accommodate the presence of the correlation hole.
Third, components associated with
$\ell > 0$, $m > 0$, and $n \ge 0$ (or $\ell > 0$, $m \ge 0$, and $n > 0$)
include electron-electron-nucleus terms,
effectively coupling the electron-electron and electron-nucleus terms previously mentioned.
These three-body terms facilitate an even more accurate representation of the correlation hole.

In the present study, we rely on the parametrization of the correlation factor, optimized within
a VMC framework at the single-determinant level, as reported in Refs.~\onlinecite{SchMos-JCP-90,Galek_2006}.

\subsection{Integral Evaluation}
\label{sec:integral_theory}
In this section, we focus on the computation of integrals necessary for TC calculations.
Besides the standard one- and two-electron integrals, one must also compute the following additional set of two-electron integrals:
\begin{equation}
	K_{r s}^{p q}
	\equiv -\mel*{\lphi{p} \lphi{q}}{\hat{K}_{12}}{\rphi{r} \rphi{s}},
	\label{eq:2e_integ_K}
\end{equation}
where $\hat{K}_{12}$ is defined in Eq.~\eqref{eq:Kop_def}.
Additionally, three-electron integrals, as defined in Eq.~\eqref{eq:3e_integ}, must be computed.

The two-electron integrals are computed in the atomic orbital basis $\{\chi_{\mu}\}$, as follows:
\begin{equation}
\begin{aligned}
	K_{\eta \zeta}^{\mu \nu}
	&\equiv -\mel{\chi_{\mu} \, \chi_{\nu}}{\hat{K}_{12}}{\chi_{\eta} \, \chi_{\zeta}}.
	\label{eq:2e_integ_K_AO}
\end{aligned}
\end{equation}
These integrals are subsequently transformed into the orbital bases to obtain Eq.~\eqref{eq:2e_integ_K}.

By exploiting the symmetry of the operator $\hat{K}_{12}$ with respect to electron exchange ($\hat{K}_{12}=\hat{K}_{21}$) and employing
integration by parts, we obtain
\begin{equation}
\begin{aligned}
	K_{\lambda \sigma}^{\mu \nu}
	&=
	\mathcal{K}_{\lambda \sigma}^{\mu \nu} + \mathcal{K}_{\sigma \lambda}^{\nu \mu}
\end{aligned}
\end{equation}
where we introduce
\begin{equation}
\begin{split}
	\mathcal{K}_{\lambda \sigma}^{\mu \nu}
	&=
	\frac{1}{2} \int \dd\br{} \Big[
		\chi_{\mu}(\br{}) \grad \chi_{\lambda}(\br{})
		- \grad \chi_{\mu}(\br{}) \chi_{\lambda}(\br{})
	\Big] \cdot \bG_{\sigma}^{\nu}(\br{}) \\
	&+
	\frac{1}{2} \int \dd\br{} \chi_{\mu}(\br{}) \chi_{\lambda}(\br{}) J_{\sigma}^{\nu}(\br{}).
\end{split}
	\label{eq:Kinteg_def}
\end{equation}
The three-dimensional integrals $\bG_{\sigma}^{\nu}$ and the scalar integrals $J_{\sigma}^{\nu}$
are defined by
\begin{subequations}
\begin{align}
        \label{eq:Ginteg_def}
        \bG_{\sigma}^{\nu}(\br{1})
        &\equiv \int \dd\br{2} \chi_{\nu}(\br{2}) \chi_{\sigma}(\br{2}) \grad_1 u_{12}, \\
        \label{eq:Jinteg_def}
        J_{\sigma}^{\nu}(\br{1})
        &\equiv \int \dd\br{2} \chi_{\nu}(\br{2}) \chi_{\sigma}(\br{2}) \qty[\grad_1 u_{12}]^2.
\end{align}
\end{subequations}

The three-electron integrals \eqref{eq:3e_integ} are computed directly in the orbital basis and are given by
\begin{equation}
\begin{split}
	L_{stu}^{pqr}
	=
	\int \dd\br{} \Big[
		& \lphi{p}(\br{}) \rphi{s}(\br{}) \bG_{t}^{q}(\br{}) \cdot \bG_{u}^{r}(\br{}) \\
		+
		& \lphi{q}(\br{}) \rphi{t}(\br{}) \bG_{s}^{p}(\br{}) \cdot \bG_{u}^{r}(\br{}) \\
		+
		& \lphi{r}(\br{}) \rphi{u}(\br{}) \bG_{s}^{p}(\br{}) \cdot \bG_{t}^{q}(\br{})
	\Big].
	\label{eq:Linteg_def}
\end{split}
\end{equation}
Note that $\bG_{q}^{p}(\br{})$ corresponds to the potential \eqref{eq:Ginteg_def} in the orbital basis.

As the integrals previously defined cannot be obtained in closed form, except for specific forms of correlation factors,
\cite{TenNo-CPL-00-b,Gin-JCP-21}
we compute numerically these integrals using Becke's quadrature grid. \cite{Bec-JCP-88b,MurKno-JCP-96}
Nevertheless, as the integrands in Eqs.~\eqref{eq:Kinteg_def} and \eqref{eq:Linteg_def} are much smoother
than those appearing in the definition of the potentials in Eqs.~\eqref{eq:Ginteg_def} and \eqref{eq:Jinteg_def},
we employ two distinct grids: one for evaluating integrals over
$\br{2}$ [as in Eqs.~\eqref{eq:Ginteg_def} and \eqref{eq:Jinteg_def}] which is denser than
the one used for evaluating integrals over $\br{}$ [as in Eqs.~\eqref{eq:Kinteg_def} and \eqref{eq:Linteg_def}].
This approach allows for an efficient computation of two- and three-electron integrals via
dense matrix-matrix multiplications using BLAS routines.

\subsection{Computational cost}
\label{sec:comp_cost}

Here, we discuss some considerations regarding the computational overhead associated with a TC-SCI calculation compared to its Hermitian version.

Regarding strictly the SCI part, the dominant computational costs stem from the evaluation of the PT2 correction and the
Davidson diagonalization.
These steps are limited by two operations:
i) applying the Slater-Condon rules between determinants
(\textit{i.e.}~finding the particle-hole excitation operators together with the fermionic phase factor)
and ii) accessing and combining the integrals needed for the computation of matrix elements.
Thanks to the normal-ordering approximation presented in Sec.~\ref{sec:nol_theory}, $\Htc$ remains a two-body operator.
Therefore, the Slater-Condon rules are unchanged and can be very efficiently computed. \cite{SceGin-arXiv-13}
In the present implementation, the left and right eigenvectors are computed separately, leading to
an overall increase of the computational time by a factor of 2. 

Except for the nonsymmetric Davidson diagonalization for large determinant spaces,
the computation of the two- and three-electron integrals is the main computational bottleneck of the TC-CIPSI approach.
To avoid storing these integrals, which would require $\order*{N^6}$ storage
(where $N$ is the number of basis functions), they are computed on the fly and contracted to form the operators defined in
Eqs.~\eqref{eq:L0}, \eqref{eq:L1}, and \eqref{eq:L2}. $\hat{L}^{(0)}$, $\hat{L}^{(1)}$, and $\hat{L}^{(2)}$ require
$O^3$, $O^2N^2$, and $ON^4$ integrals, respectively, while the construction of the TC Fock operator involves $O^2N^2$
integrals [see Eq.~\eqref{eq:tc_fock}], where $O$ the number of occupied orbitals.
To compute $L_{stu}^{pqr}$, one first evaluates the potentials $\bG_{p}^{q}(\br{})$ on each grid point for each pair of
orbitals, which necessitates $K N^2$ storage (where $K$ is the number of grid points). Then, these values are employed
in Eqs.~\eqref{eq:L0}, \eqref{eq:L1}, and \eqref{eq:L2} to compute the required integral batches.

\section{Results and discussion}
\label{sec:results}

\subsection{Computational details}
\label{sec:comp_det}
The calculations reported in this manuscript are performed with \textsc{quantum package}. \cite{QP2-JCTC-19}
In the subsequent calculations, we employ either the 6-31G split-valence basis sets \cite{Ditchfield_1971,Hehre_1972,Binkley_1980}
or Dunning's cc-pVXZ and cc-pCVXZ basis sets, where X denotes D, T, Q, and 5. \cite{Dun-JCP-89,WooDun-JCP-95}
The molecular geometries were extracted from the \textsc{quest} database \cite{LooSceJac-JPCL-20,VerSceCafLipBogJacLoo-WIRE-21,MarLoo-arXiv-24}
or experimental data, \cite{HubHer-BOOK-79} and are provided in the \SupInf.
Unless otherwise stated, calculations are performed in the frozen-core approximation.

The Jastrow parameters were obtained from two sources: Ref.~\onlinecite{SchMos-JCP-90} and Ref.~\onlinecite{Galek_2006}.
Specifically, for atoms, we utilized the parameters outlined in Table VI of Ref.~\onlinecite{SchMos-JCP-90}, whereas for molecules,
the Jastrow parameters were sourced from Table 2 of Ref.~\onlinecite{Galek_2006}.
Mean-field calculations on closed-shell systems are performed in the restricted formalism while the restricted open-shell
formalism is employed for open-shell systems. \cite{AmmSceGin-JCTC-23}
For the latter case, an average of integrals corresponding to spin-up and spin-down densities is employed in the normal-ordering procedure.
In the calculations presented herein, we use a grid with 30 radial and 50 angular points for each atom over $\br{}$, and 70 radial and 266 angular points over $\br{2}$.
These grid configurations ensure the stability of energy calculations to within at least \SI{0.1}{\milli\hartree}.
All calculations were carried out on a single dual-socket node equipped with 36 Intel Skylake 6140 cores, each running at 2.3 GHz.

\subsection{Weights of the Slater determinants}
\label{sec:conv_weight}
We begin this study by comparing the weight $c_I^2$ of a given Slater determinant $I$ (built with HF orbitals)
in the ground-state FCI wave function,
and the pseudo-weight $\abs{\tilde{c}_I c_I}$ of the same determinant in the ground-state TC-FCI wave function.
As a first example, we consider the \ce{H2O} molecule in the 6-31G basis set as this relatively small basis set enables the calculation of both the FCI and TC-FCI wave functions.

\begin{figure*}
	\includegraphics[width=0.49\textwidth]{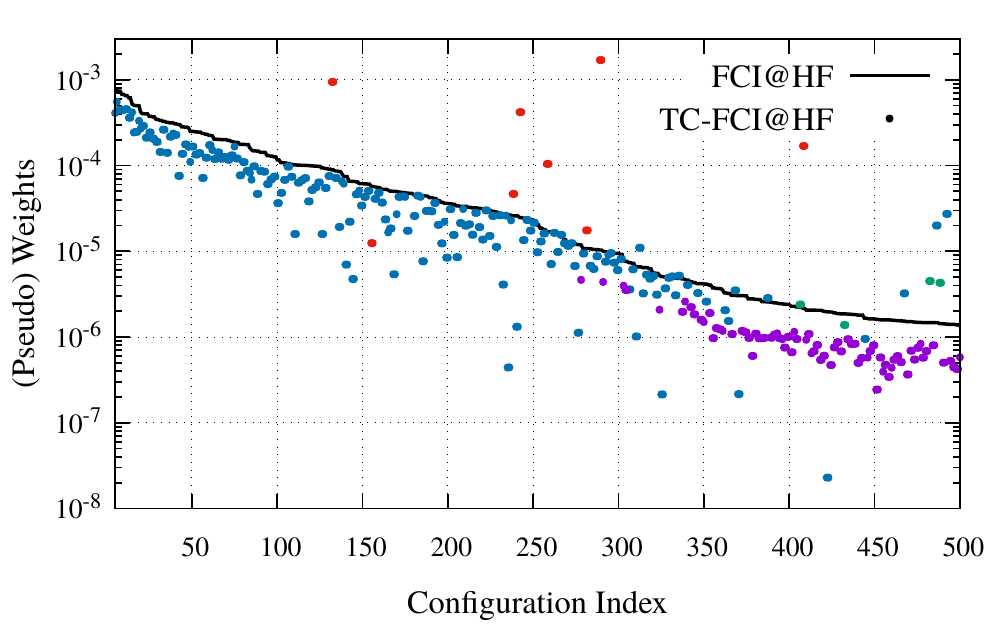}
	\includegraphics[width=0.49\textwidth]{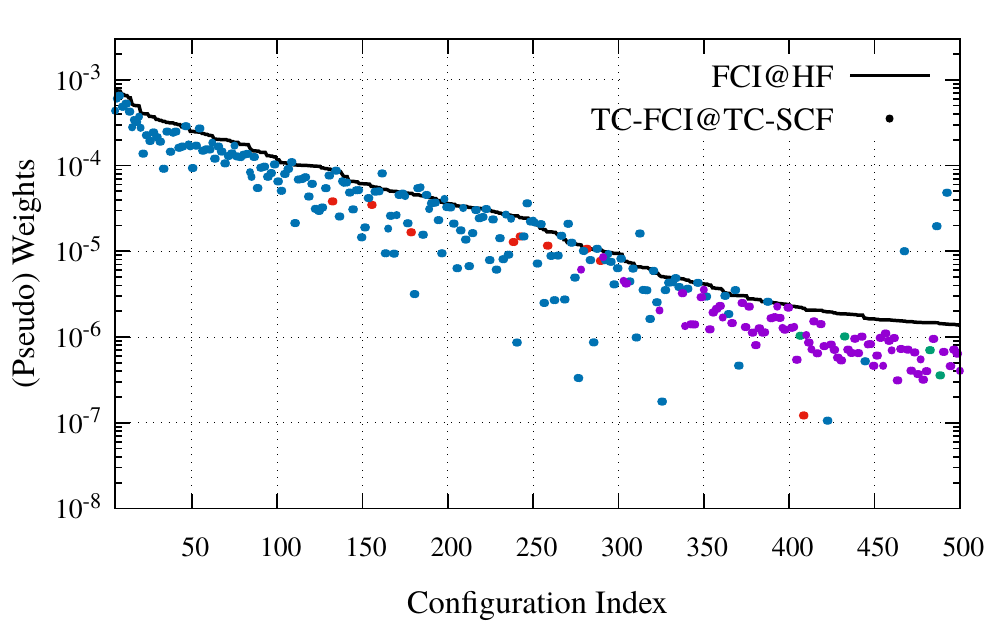}
	\caption{
	Weights ($c_I^2$) of the FCI wave function (solid black line) and pseudo-weights ($\abs{\tilde{c}_I c_I}$)
	of the TC-FCI wave function (colored markers) for \ce{H2O} in the 6-31G basis set.
	The excitation degree of each determinant with respect to the mean-field reference determinant is
	indicated by the following color code: red, blue, green, and purple for single, double,
	triple, and quadruple excitations, respectively.
	The left panel displays the weight distribution of the determinants sorted by $c_I^2$,
	using HF orbitals to construct the determinants.
	The right panel reports similar quantities but TC-SCF orbitals are employed for constructing the TC-FCI wave function.
	}
	\label{fig:H2O_WF}
\end{figure*}

We report in the left panel of Fig.~\ref{fig:H2O_WF} the FCI weights $c_I^2$ associated with the 500 most important Slater determinants sorted by decreasing weights,
as well as the corresponding TC-FCI pseudo-weights $\abs{\tilde{c}_I c_I}$.
Additionally, the level of excitation with respect to the reference determinant is indicated by different colors.
An important observation is that the pseudo-weights do not decrease monotonically, unlike the weights. This indicates that when using the TC Hamiltonian, the determinants are not selected in the same order as with the standard Hamiltonian when using the CIPSI algorithm.
Also, the pseudo-weights are globally smaller than the FCI weights, except for the single excitations
which are typically much larger in the TC wave function. The latter observation can be explained by the fact that while HF orbitals
fulfill the Brillouin theorem for the standard Hamiltonian, it is no longer the case for the TC Hamiltonian
as the effective interaction is no longer the bare Coulomb repulsion.

We report, in the right panel of Fig.~\ref{fig:H2O_WF}, the weight distribution of the TC
wave function using now TC-SCF orbitals,
which therefore fulfills both the left and right Brillouin conditions [see Eqs.~\eqref{eq:brillouin_left} and \eqref{eq:brillouin_right}].
Consequently, the pseudo-weights of the single excitations are significantly reduced.
For the sake of completeness, we should mention that in the latter case, as the orbitals are different,
the determinants in the FCI and TC-FCI wave functions are no longer the same
but here we simply match a given determinant from the two wave functions with its orbital occupation,
assuming that the physical meaning of the orbitals weakly changes with the TC-SCF optimization (which is clearly the case here).

More quantitatively, there are \num{2590} determinants with a weight larger than \num{d-7}
in the FCI wave function, while there are only \num{1998} and \num{1704} determinants
with a pseudo-weight larger than \num{d-7} in the TC wave function using the HF and TC-SCF orbitals, respectively.
This shows that the correlation factor has the effect of lowering the number of determinants that significantly contribute to the energy. This is the first evidence of the compactification of the determinant expansion brought by the TC approach.

\begin{figure}
	\includegraphics[width=1.00\linewidth]{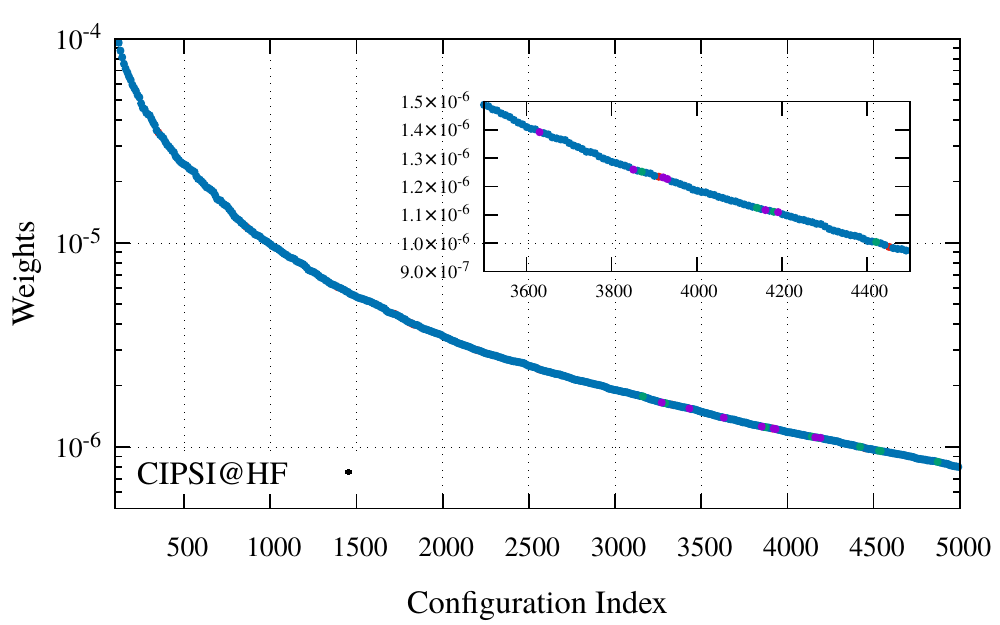}
	\includegraphics[width=1.00\linewidth]{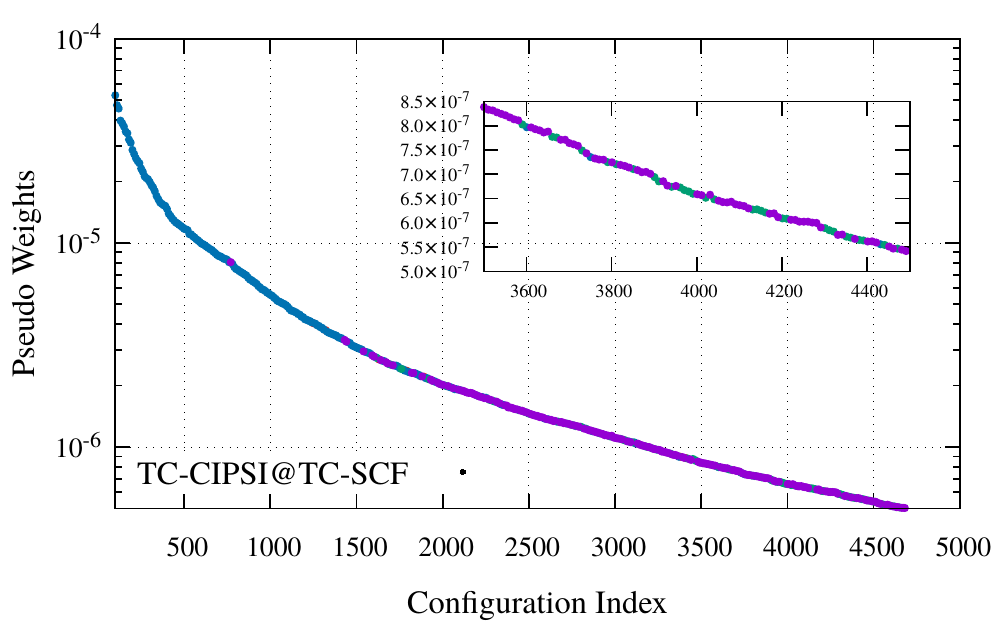}
	\caption{
	Weights ($c_I^2$) or pseudo-weights ($\abs{\tilde{c}_I c_I}$) of the first 5000 determinants composing the CIPSI (left) and TC-CIPSI (right) wave functions, sorted with respect to their corresponding (pseudo) weights, for \ce{H2O} in the cc-pVTZ basis set.
	For the sake of readability, one point out of 10 is represented.
	The excitation degree of each determinant with respect to the mean-field reference determinant is
	indicated by the following color code: red, blue, green, and purple for single, double,
	triple, and quadruple excitations, respectively.
	}
	\label{fig:H2O_VTZ}
\end{figure}

Figure \ref{fig:H2O_VTZ} shows similar quantities but in a larger basis set (cc-pVTZ) and computed at the CIPSI and TC-CIPSI levels employing HF and TC-SCF orbitals, respectively.
In this case, the determinants are sorted with respect to their corresponding (pseudo) weights in their corresponding wave functions.
The same color code is used to indicate the excitation degree of the determinants.
Additionally, to study the impact of the form of the correlation factor, we performed calculations using the simple correlation factor introduced previously by some of the authors. \cite{AmmGinSce-JCTC-22} The latter form consists of a universal two-body correlation factor designed for valence electrons multiplied by atom-centered Gaussian envelopes suppressing the effect of the correlation factor in the core regions. The nuclear parameters for the Gaussian envelopes used for these calculations were taken from Ref.~\onlinecite{AmmSceGin-JCP-23}. These complementary results are reported in the \SupInf.

Three key observations can be made:
i) as above, the pseudo weights are systematically lower than the weights,
ii) triple and higher excitations enter much earlier in the TC-CIPSI expansion, which demonstrates that the correlation factor allows for a significant reduction of the weights of doubly-excited determinants, and
iii) the above effects are considerably less pronounced with the simpler correlation factor, highlighting the importance of flexible functional forms.

\subsection{Convergence of the zeroth-order energy and its extrapolation}
\label{sec:conv_extrap}
A critical aspect of the TC-SCI scheme is the convergence of the zeroth-order energy $E_{\text{TC}}^{(0)}$ based
on the criterion employed to select Slater determinants, a smooth convergence facilitating its extrapolation towards the TC-FCI limit.
Here, we would like to investigate two distinct selection procedures that evidence the impact of the correlation factor:
i) selecting determinants via the second-order contribution computed with the TC Hamiltonian
using both left and right eigenvectors [see Eq.~\eqref{eq:second_e}] (TC-CIPSI selection),
or ii) performing a standard selection based on the bare Hamiltonian (CIPSI selection).

\begin{figure}
	\includegraphics[width=\linewidth]{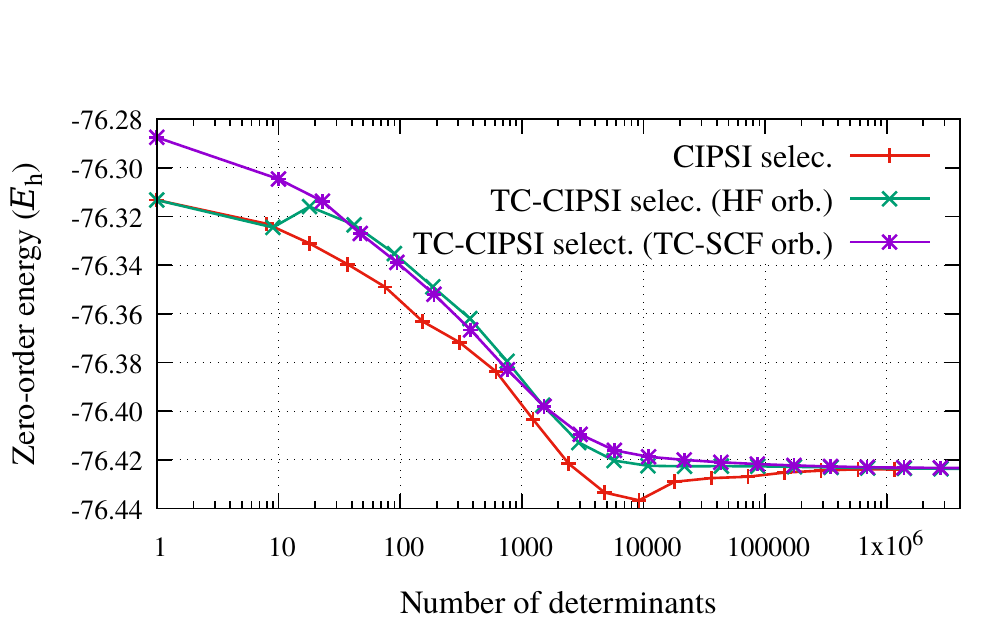}
 	\includegraphics[width=\linewidth]{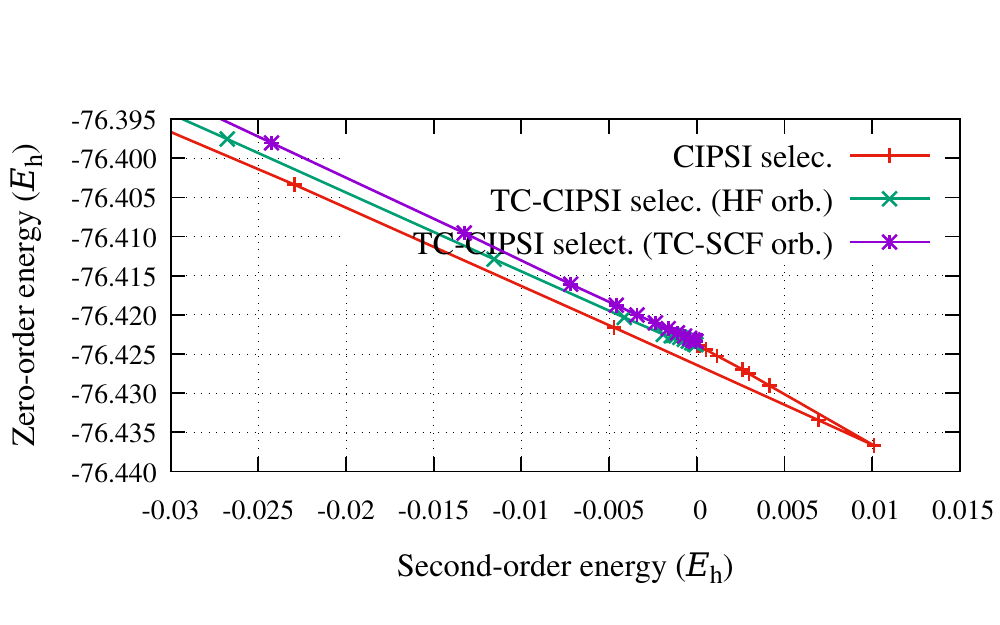}
	\includegraphics[width=\linewidth]{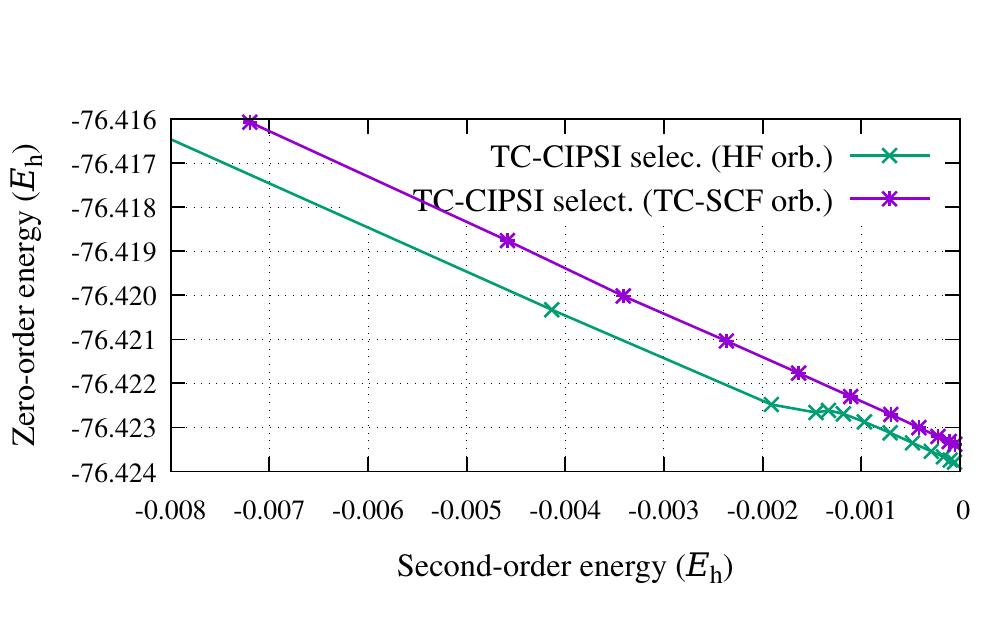}
	\caption{
        Convergence of $E_{\text{TC}}^{(0)}$ as a function of the number of selected determinants
	(top) or $E_{\text{TC}}^{(2)}$ (center and bottom) for \ce{H2O} computed in the cc-pVTZ basis
	and obtained via the CIPSI and TC-CIPSI selection procedures.
        In the case of TC-CIPSI, HF or TC-SCF orbitals are employed.
        The bottom panel corresponds to a zoom of the central panel near the origin.}
	\label{fig:h2o_conv}
\end{figure}

To illustrate this point, we rely on the same representative example, the \ce{H2O} molecule in the cc-pVTZ basis set.
We report the convergence of $E_{\text{TC}}^{(0)}$ with respect to the number of Slater determinants
in Fig.~\ref{fig:h2o_conv} with the CIPSI and TC-CIPSI selection procedures discussed above.
We observe that, starting from the HF reference determinant,
the CIPSI selection leads rapidly (approximately \num{3000} determinants)
to an energy that is below the estimated TC-FCI energy, while, for expansions larger than \num{600 000} determinants,
$E_{\text{TC}}^{(0)}$ rises to eventually reach this limiting value.
If one selects the determinants using the TC-CIPSI algorithm instead, we observe that, from about 20
determinants, the zeroth-order energy monotonically decays and remains above the TC-FCI energy.
This clearly highlights the impact of the correlation factor on the selection of the determinants.
Additionally, Fig.~\ref{fig:h2o_conv} illustrates the convergence of $E_{\text{TC}}^{(0)} $using TC-SCF orbitals, as opposed to HF orbitals, highlighting the impact of orbital optimization in the TC framework.
One can see that although the TC-SCF reference determinant is higher in energy, $E_{\text{TC}}^{(0)}$
rapidly and monotonically decay towards the TC-FCI energy.

In Fig.~\ref{fig:h2o_conv}, we also report the evolution of $E_{\text{TC}}^{(0)}$ as a function of the second-order
energy correction $E_{\text{TC}}^{(2)}$ [see Eq.~\eqref{eq:second_e_tot}] for the various cases investigated here.
This helps us to appreciate the different behaviors of $E_{\text{TC}}^{(0)}$ from the perspective of its extrapolation.
On the one hand, we observe that the erratic behavior of $E_{\text{TC}}^{(0)}$
obtained from the CIPSI selection makes it hard to extrapolate $E_{\text{TC}}^{(0)}$ to the TC-FCI energy, due to the presence of a ``turning'' point stemming from the interplay between the positive and negative contributions of $E_{\text{TC}}^{(2)}$, as mentioned in Sec.~\ref{sec:sci_theory}.
On the other hand, using the TC-CIPSI selection leads to a globally linear curve.
However, when one employs HF orbitals, we observe a sudden change of slope as $E_{\text{TC}}^{(2)} \to 0$,
leading also to a possible untrustworthy extrapolation.
However, using TC-SCF orbitals, $E_{\text{TC}}^{(0)}$ has a consistent linear behavior on a larger range of $E_{\text{TC}}^{(2)}$
allowing us to estimate the TC-FCI energy more safely.
It is worth mentioning the small difference (roughly \SI{0.4}{\milli\hartree})
in the TC-FCI limiting values obtained with HF and TC-SCF orbitals,
which is due to the normal-ordering approximation that necessarily creates a weak dependence on the orbitals.

We can then conclude that the best strategy to perform a TC-SCI calculation
is to rely on TC-SCF orbitals in combination with the TC-CIPSI algorithm for the determinant selection.
In such a way, one obtains rapidly and monotonically convergent zeroth-order energies that permit a reliable extrapolation to the TC-FCI limit.
From hereon, TC-CIPSI calculations are systematically performed with TC-SCF orbitals.

\subsection{Compactification of the CI expansion}
\label{sec:compact}

To demonstrate the disparity in convergence rate between energies computed via the CIPSI and TC-CIPSI
methods, we conduct calculations on the water molecule using increasingly large Dunning basis sets (cc-pVDZ,
cc-pVTZ, and cc-pVQZ). In order to quantify the acceleration brought by the TC approach, we primarily focus on two indicators of convergence.
The first criterion corresponds to the number of determinants required to reach a specified value of the second-order energies, $E^{(2)}$ for CIPSI and $E_{\text{TC}}^{(2)}$ for TC-CIPSI.
The second criterion involves achieving stable extrapolations towards the FCI and TC-FCI limits, respectively, through fitting functions represented as
\begin{subequations}
\begin{align}
	E^{(0)} \qty(E^{(2)}) & = a E^{(2)} + b,
	\label{eq:extrap_E}
	\\
	E_{\text{TC}}^{(0)} \qty(E_{\text{TC}}^{(2)}) & = a_{\text{TC}} E_{\text{TC}}^{(2)} + b_{\text{TC}}.
	\label{eq:extrap_ETC}
\end{align}
\end{subequations}
The data associated with the following analysis can be found in the \SupInf.

\begin{figure}
	\includegraphics[width=\linewidth]{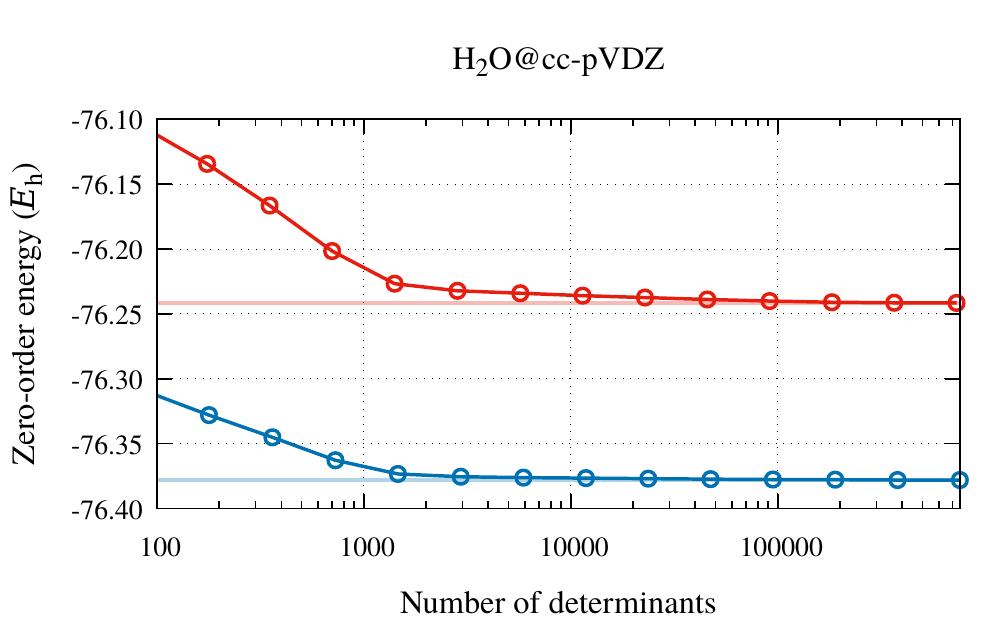}
	\includegraphics[width=\linewidth]{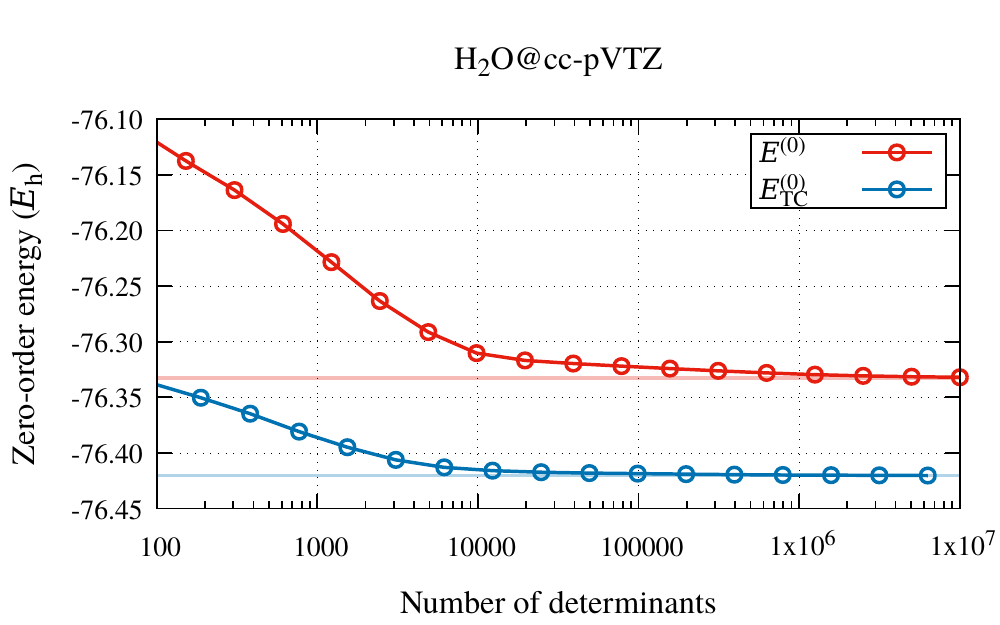}
	\includegraphics[width=\linewidth]{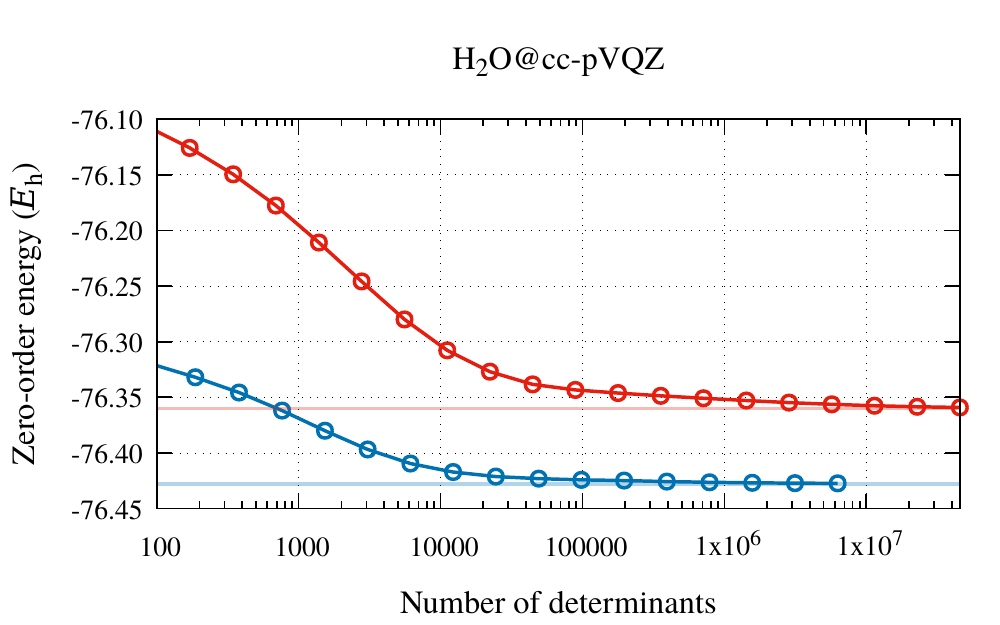}
	\caption{
	Zeroth-order energies, $E^{(0)}$ (red) and $E_{\text{TC}}^{(0)}$ (blue),
	as functions of the number of selected determinants for \ce{H2O} in the
	cc-pVDZ (top), cc-pVTZ (center), and cc-pVQZ (bottom) basis.
	The shaded horizontal lines represent the extrapolated FCI (red) and TC-FCI (blue) energies.}
	\label{fig:H2O_FCcalc_conve}
\end{figure}

\begin{figure}
	\includegraphics[width=\linewidth]{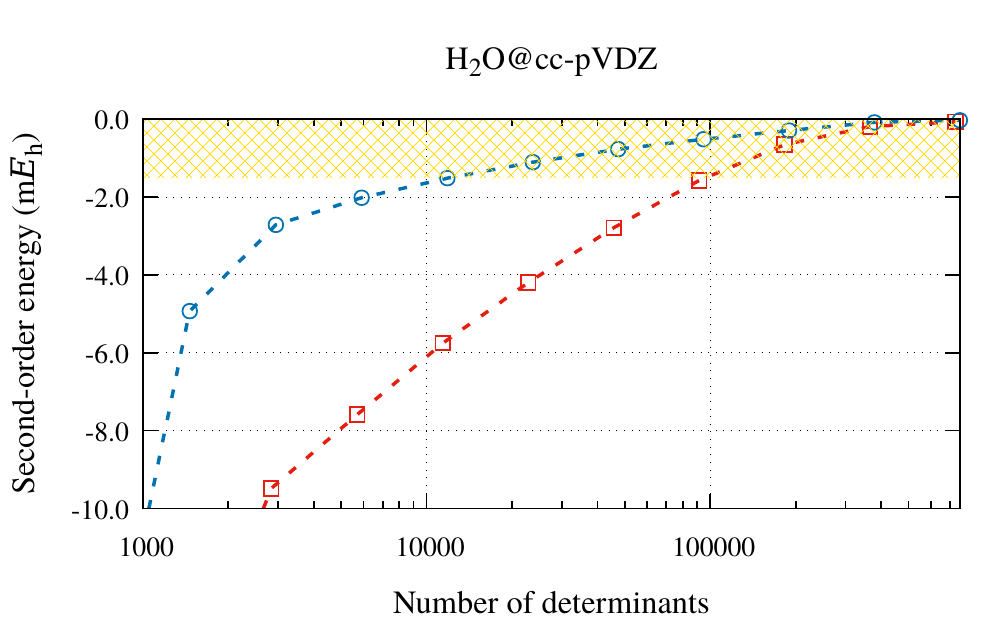}
	\includegraphics[width=\linewidth]{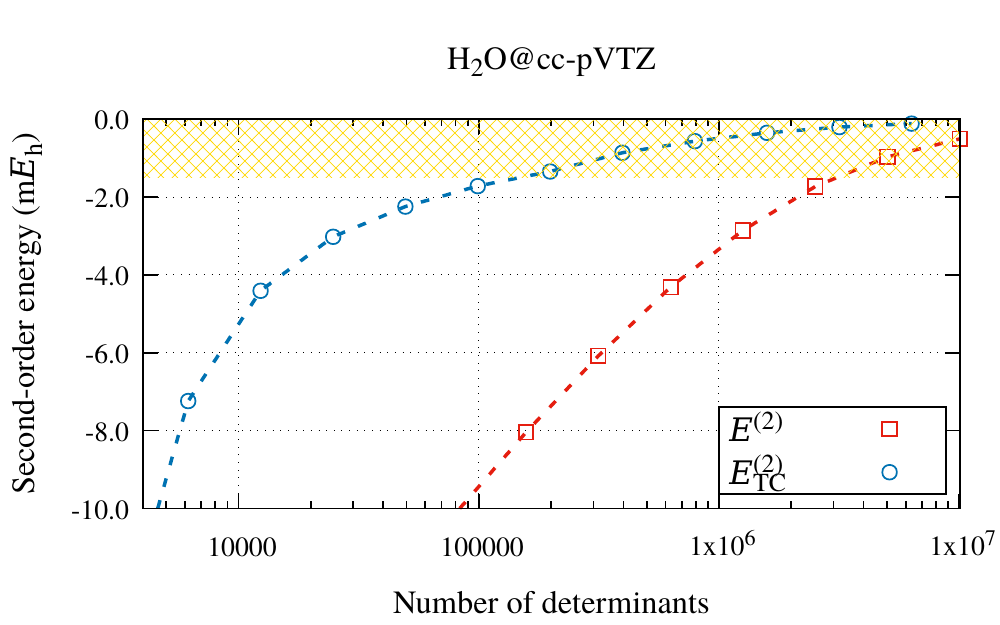}
	\includegraphics[width=\linewidth]{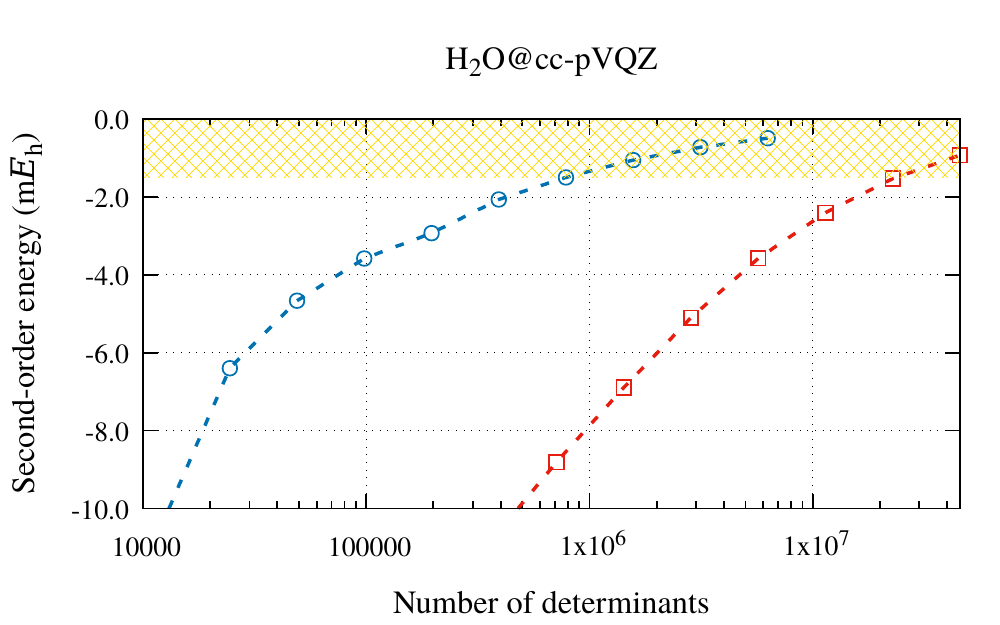}
	\caption{
	Second-order perturbative energies, $E^{(2)}$ (red) and $E_{\text{TC}}^{(2)}$ (blue), as functions of the number of selected determinants
	for \ce{H2O} in the cc-pVDZ (top), cc-pVTZ (center), and cc-pVQZ (bottom) basis.
	The shaded yellow region corresponds to \SI{1.5}{\milli\hartree} accuracy.
	}
	\label{fig:H2O_FCcalc_pt2}
\end{figure}

The convergence behavior of the zeroth-order energies, $E^{(0)}$ and $E_{\text{TC}}^{(0)}$, are depicted in Fig.~\ref{fig:H2O_FCcalc_conve}. The corresponding extrapolated FCI and TC-FCI energies, $E_{\text{FCI}}$ and $E_{\text{TC-FCI}}$, are presented by shaded lines.
These plots illustrate that, as the number of selected determinants increases, $E_{\text{TC}}^{(0)}$ converges
towards $E_{\text{TC-FCI}}$ at a notably faster rate compared to the convergence of $E^{(0)}$ towards $E_{\text{FCI}}$.
A similar trend is observed for the second-order perturbative energies, as shown in Fig.~\ref{fig:H2O_FCcalc_pt2}, where we report the evolution of $E^{(2)}$ and $E_{\text{TC}}^{(2)}$ as functions of the number of determinants.

More quantitatively, achieving chemical accuracy, which corresponds to values of $E_{\text{PT2}}$ below \SI{1.5}{\milli\hartree}, requires
approximately \num{96 883}, \num{3 105 054}, and \num{23 609 437} determinants in the cc-pVDZ, cc-pVTZ, and cc-pVQZ basis sets,
respectively. Conversely, in the TC case, reaching $E_{\text{PT2}}^{\text{TC}} \approx \SI{1.5}{\milli\hartree}$ occurs much
earlier, with approximately \num{12 195}, \num{148 821}, and \num{781 281} determinants in these bases, indicating a compactification of the CI expansion by factors of approximately $8$, $21$, and $30$, respectively.

\begin{figure}
	\includegraphics[width=\linewidth]{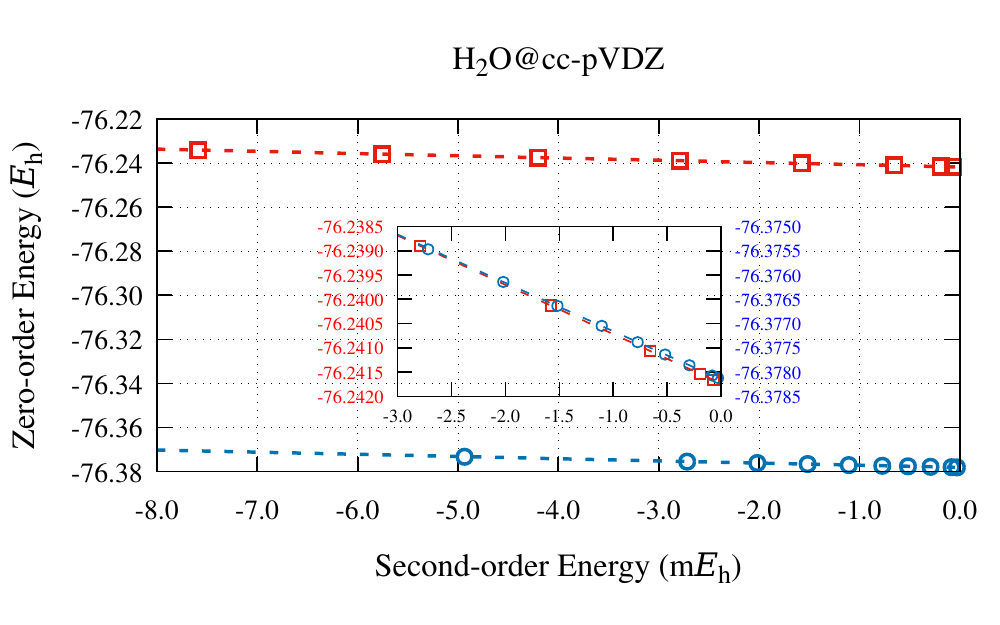}
	\includegraphics[width=\linewidth]{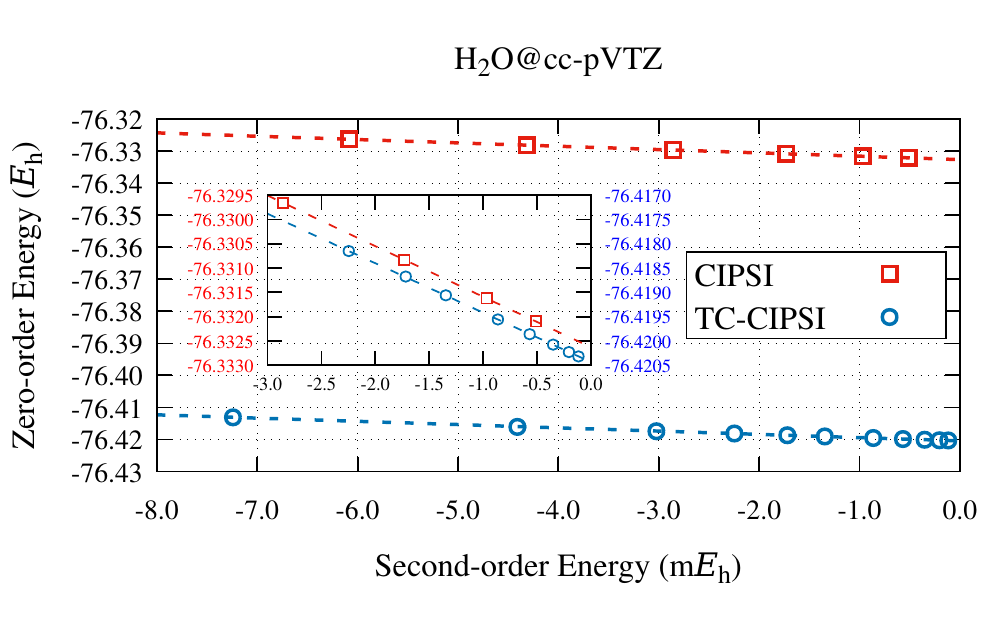}
	\includegraphics[width=\linewidth]{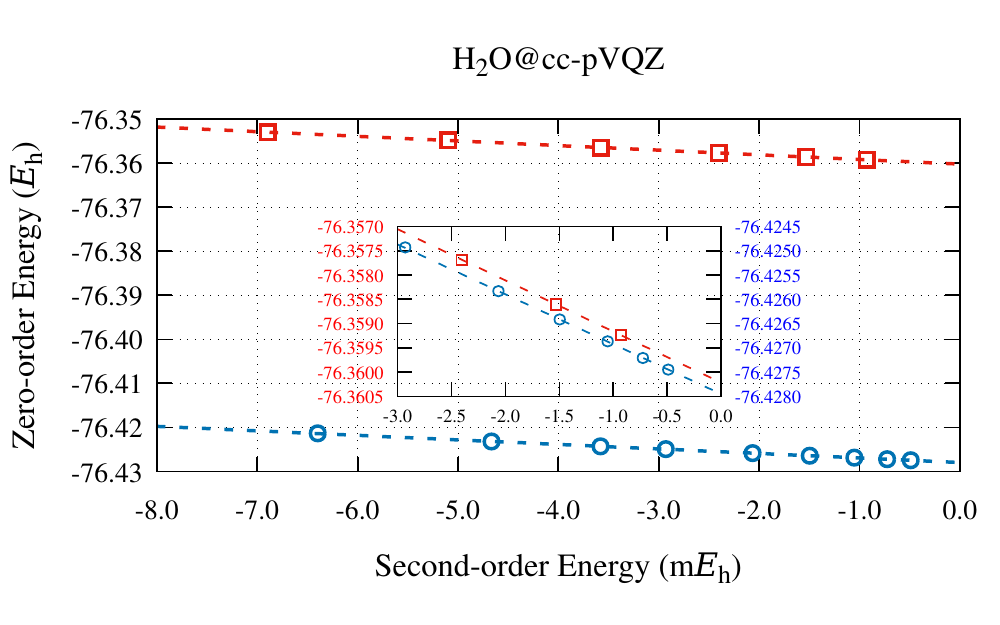}
	\caption{
	Linear extrapolations of $E^{(0)}$ (red) and $E_{\text{TC}}^{(0)}$ (blue)
	towards the FCI and TC-FCI limiting energies for \ce{H2O}
	in the cc-pVDZ (top), cc-pVTZ (center), and cc-pVQZ (bottom) basis.
	}
	\label{fig:H2O_FCcalc_extrap}
\end{figure}

In Fig.~\ref{fig:H2O_FCcalc_extrap}, we depict the evolution of $E^{(0)}$ as a function of $E^{(2)}$, as well as $E_{\text{TC}}^{(0)}$ as a function of $E_{\text{TC}}^{(2)}$. We perform a linear fit of these data using
Eqs.~\eqref{eq:extrap_E} and \eqref{eq:extrap_ETC} to target the FCI and the TC-FCI energies, respectively.
Across all three basis sets, both CIPSI and TC-CIPSI demonstrate stable and smooth linear extrapolations.
However, upon closer examination, we observe that the TC-CIPSI extrapolation converges significantly faster.

\begin{figure}
	\includegraphics[width=\linewidth]{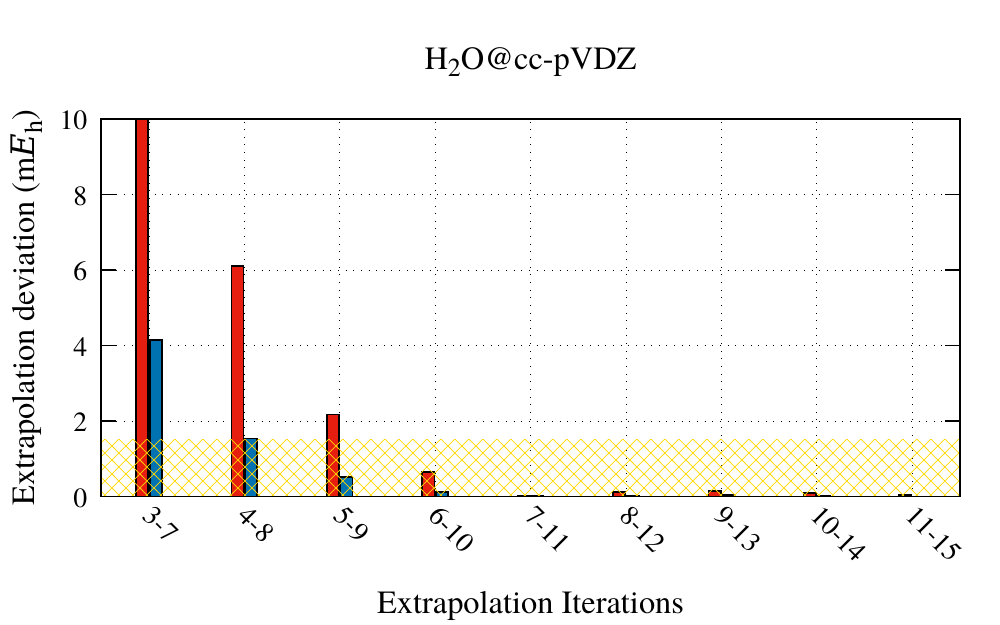}
	\includegraphics[width=\linewidth]{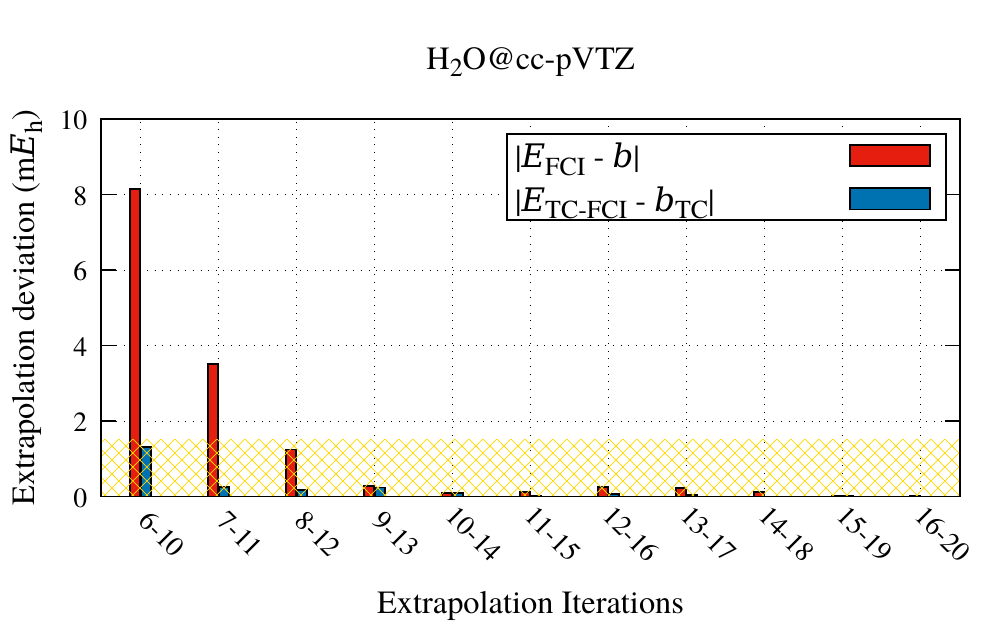}
	\includegraphics[width=\linewidth]{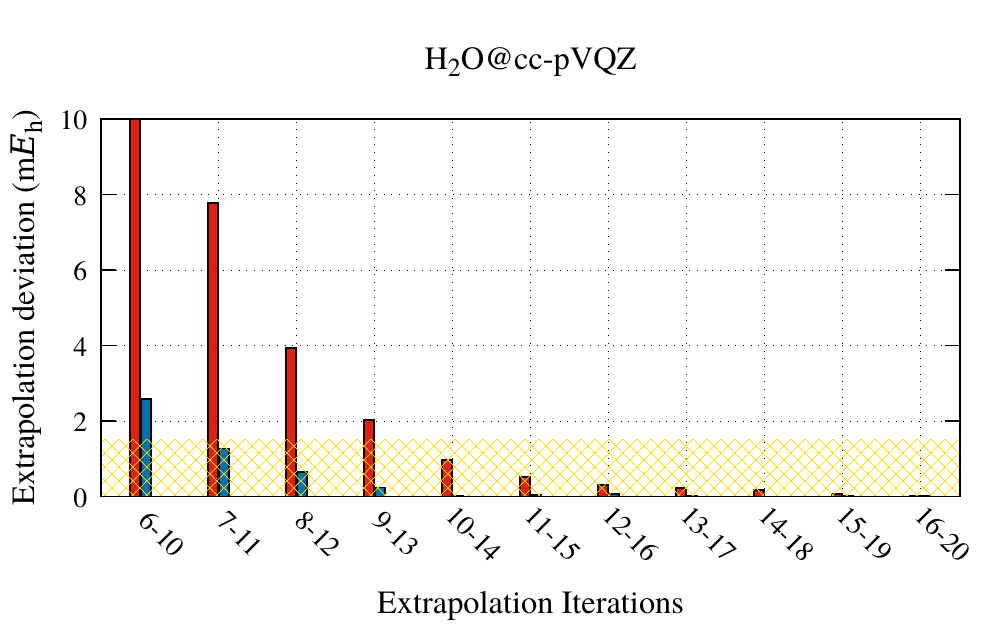}
	\caption{
	Error in extrapolated energies as additional iterations (or selected determinants) are performed.
	The shaded yellow region corresponds to \SI{1.5}{\milli\hartree} accuracy.
	}
	\label{fig:H2O_FCcalc_compextrap}
\end{figure}

In Fig.~\ref{fig:H2O_FCcalc_compextrap}, we assess the error on the extrapolated energies as additional iterations are performed
(or, equivalently, more determinants are added) with each extrapolation conducted using five consecutive iterations
of the CIPSI or TC-CIPSI algorithm.
The comparison is carried out for cc-pVDZ, cc-pVTZ, and cc-pVQZ.
We establish convergence over extrapolation when performing an additional CIPSI (TC-CIPSI)
iteration no longer alters the estimated FCI (TC-FCI) energy by more than \SI{1.5}{\milli\hartree}.
Under this criterion, the conventional CIPSI extrapolation necessitates approximately \num{2 834},
\num{9 821}, and \num{22 248} determinants to achieve convergence, while TC-CIPSI requires only
\num{729}, \num{3 085}, and \num{3 063} in the cc-pVDZ, cc-pVTZ, and cc-pVQZ bases, respectively.

\begin{figure}[]
	\includegraphics[width=\linewidth]{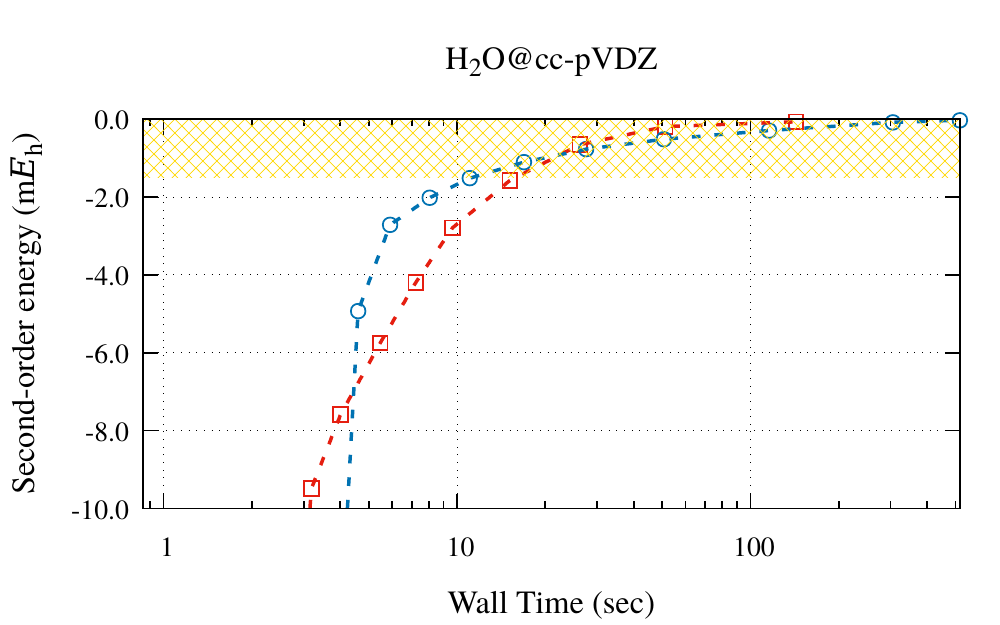}
	\includegraphics[width=\linewidth]{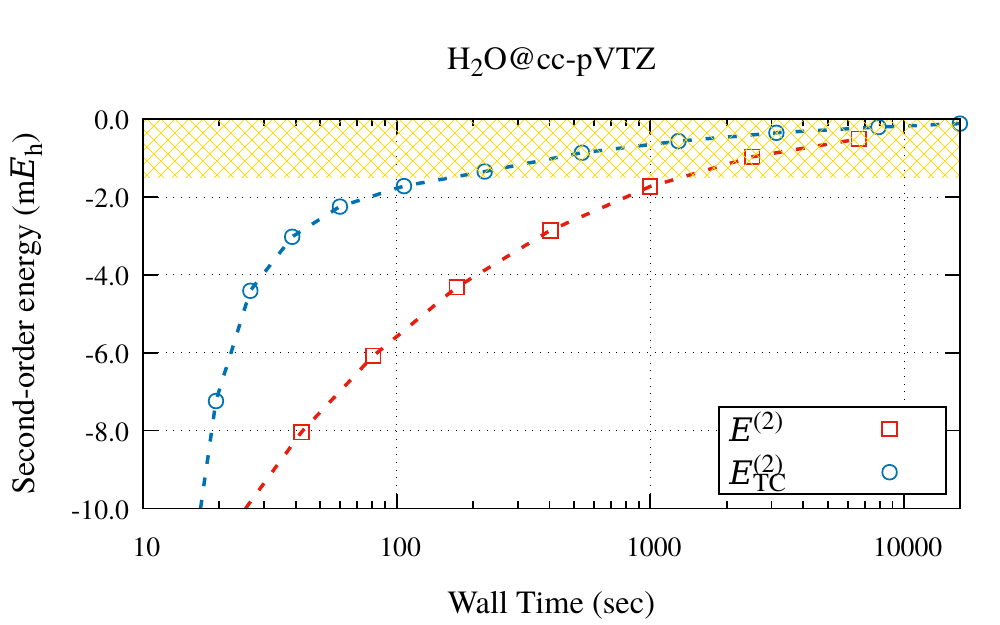}
	\includegraphics[width=\linewidth]{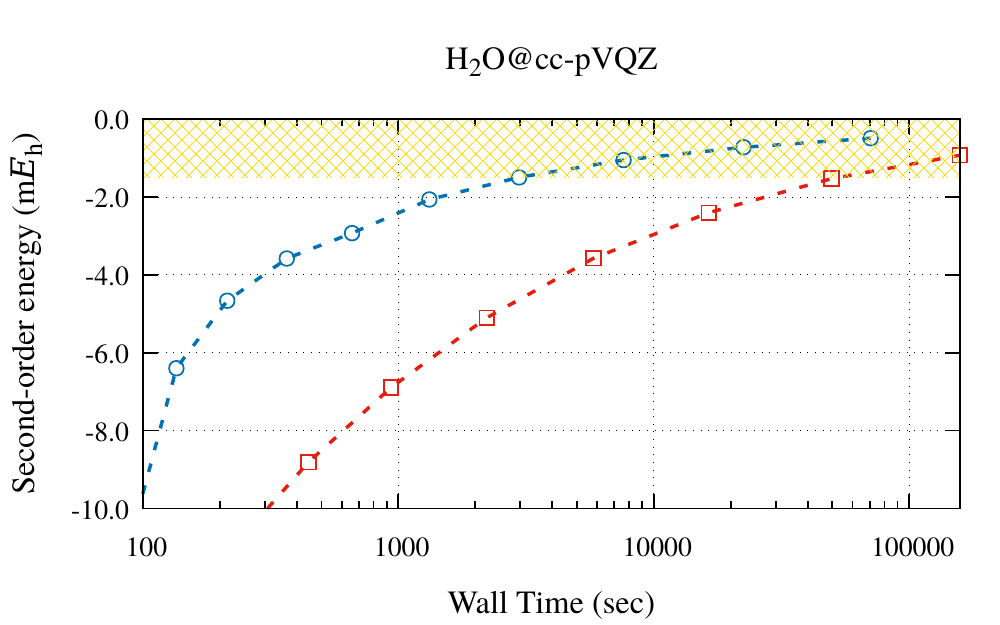}
	\caption{
	Second-order perturbative energies, $E^{(2)}$ (red) and $E_{\text{TC}}^{(2)}$ (blue), as functions of the elapsed wall time (in seconds) 	for \ce{H2O} in the cc-pVDZ (top), cc-pVTZ (center), and cc-pVQZ (bottom) basis.
	The shaded yellow region corresponds to \SI{1.5}{\milli\hartree} accuracy.
        }
        \label{fig:H2O_FCcalc_wt}
\end{figure}

Next, we compare the computational costs of CIPSI and TC-CIPSI calculations.
Figure \ref{fig:H2O_FCcalc_wt} illustrates the convergence rates of the second-order energies as a function of the elapsed
wall time (in seconds) for the same collection of basis sets (cc-pVDZ, cc-pVTZ, and cc-pVQZ).
In the smallest cc-pVDZ basis set, TC-CIPSI initially appears slower than CIPSI due to the time required for preparing the normal-ordered intermediates, as defined in Eqs.~\eqref{eq:L0}, \eqref{eq:L1}, and \eqref{eq:L2}.
However, across all basis sets, TC-CISPI demonstrates a faster convergence to chemical accuracy compared to CIPSI.
Specifically, CIPSI takes 15 seconds, 22 minutes, and 14.6 hours to reach a chemically-accurate energy in cc-pVDZ, cc-pVTZ, and cc-pVQZ, respectively.
These times are reduced in the case of TC-CIPSI, requiring only 11 seconds, 3 minutes, and 49 minutes, respectively.
Consequently, the overall gain in wall time for chemically-accurate energies using TC-CIPSI compared to CIPSI is $1.4$, $8$, and $18$, respectively.

Similar calculations were conducted for \ce{Li2}, \ce{Be2}, \ce{NH3}, \ce{CH4}, and \ce{H2CO} molecules
(the corresponding figures and tables are provided as \SupInf), yielding comparable findings.
The convergence behavior across different basis sets and the comparative analysis
of the zeroth-order, second-order, and extrapolated energies between the
CIPSI and TC-CIPSI methods consistently yielded similar results for these molecules.

\subsection{Total Energies}
\label{sec:total_energies}

Estimating the total nonrelativistic electronic energy of atomic and molecular systems presents significant challenges
for standard CI techniques due to the need for very large basis sets to achieve chemical accuracy.
However, at the TC-CI level, it is possible to reduce considerably the basis set error.
We perform TC-CIPSI calculations without the frozen-core approximation using the cc-pVDZ, cc-pVTZ, and cc-pVQZ basis sets for the
neutral atoms from $Z=2$ to $Z=10$ (Table \ref{tab:ETC_total_atom}) and for the following molecules reported in Table \ref{tab:ETC_total_molec}: \ce{Li2}, \ce{Be2}, \ce{H2O}, \ce{NH3},
\ce{CH4}, and \ce{H2CO}.
The TC-FCI energies are obtained through extrapolation of the TC-CIPSI energies, ensuring a convergence
of at least four digits (\SI{0.1}{\milli\hartree}).
It is worth noting that we opted for calculations in cc-pVXZ rather than cc-pCVXZ basis sets, as we observed that with
sufficiently large basis sets (typically when $\text{X} > 3$), the TC-FCI energies obtained in cc-pVXZ and cc-pCVXZ are very close thanks to the effect of the correlation factor.
For instance, the energy difference between cc-pVQZ and cc-pCVQZ for systems like \ce{Be}, \ce{Ne}, and \ce{H2O} is
only \SI{0.1}{\milli\hartree}, \SI{0.6}{\milli\hartree}, and \SI{0.2}{\milli\hartree}, respectively.
Such differences are significantly smaller than those observed at the standard FCI level, often by two or three orders of magnitude.

The exact total energies for atomic systems, as estimated in Ref.~\onlinecite{ChaGwaDavParFro-PRA-93}, are also presented
in Table \ref{tab:ETC_total_atom}. A remarkable agreement between the TC-FCI and reference energies is observed, particularly
with the cc-pVQZ basis set. The mean absolute errors are \SI{32.5}{\milli\hartree}, \SI{9.2}{\milli\hartree},
and \SI{0.9}{\milli\hartree} in cc-pVDZ, cc-pVTZ, and cc-pVQZ, respectively. It is noteworthy that our results closely
align with those obtained in Ref.~\onlinecite{CohLuoGutDobTewAla-JCP-19} using the TC-FCIQMC algorithm, where three-electron
integrals are treated exactly, that is, without normal-ordering. This suggests that the normal-ordering approximation
outlined in Sec.~\ref{sec:nol_theory} introduces minimal bias, even in the case of open-shell systems.

\begin{figure}[]
	\includegraphics[width=\linewidth]{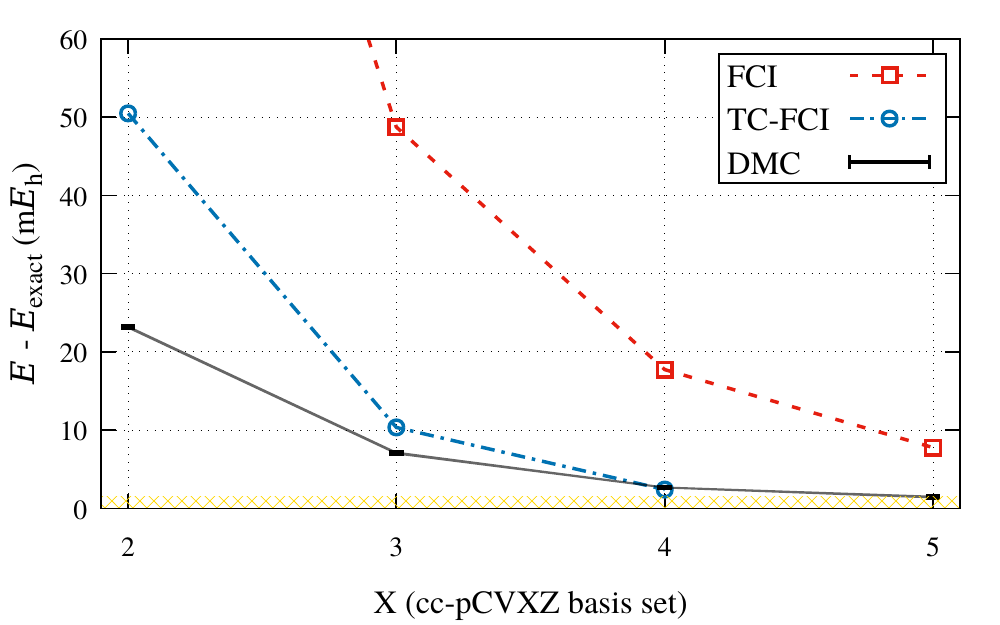}
	\caption{
	Error (in \si{\milli\hartree}) in the total energy of \ce{H2O} computed at the
	FCI, TC-FCI, and FN-DMC levels with the cc-pCVXZ family of basis sets
	with respect to the estimated exact energy (\SI{-76.4389}{\hartree}).
	The shaded yellow region corresponds to \SI{1.5}{\milli\hartree} accuracy.
	}
	\label{fig:Etotal_water}
\end{figure}

Finding exact estimates of nonrelativistic total energies for molecular systems in the literature is
significantly more challenging, with very few exceptions.
All-electron fixed-node diffusion Monte Carlo (FN-DMC) calculations for the nonrelativistic ground-state energy
of \ce{H2O} has been reported in Ref.~\onlinecite{CafAplGinSce-JCP-16} using the cc-pCVXZ basis set family.
Consequently, we conducted additional calculations on the water molecule using these basis sets. The comparison between
FCI, \cite{AlmoraDiaz_2014} TC-FCI, and FN-DMC~\cite{CafAplGinSce-JCP-16} calculations is presented in Table \ref{tab:Etotal_water} and visualized in
Fig.~\ref{fig:Etotal_water}, with the reference exact energy considered as \SI{-76.4389}{\hartree}. \cite{Klo-MP-01}
It is evident that while the standard FCI energy converges very slowly, the TC-FCI energy is found to be very close to the FN-DMC
energy in cc-pVTZ and nearly identical in cc-pVQZ.

\begin{table*}
\caption{
Total energies (in \si{\hartree}) obtained from TC-FCI calculations (without frozen-core approximation) in the cc-pVDZ, cc-pVTZ, and
cc-pVQZ basis sets for the neutral atoms from $Z=2$ to $Z=10$.
For the sake of comparison, estimated exact energies are also provided.
}
\label{tab:ETC_total_atom}
\begin{ruledtabular}
\begin{tabular}{cccccccccc}
Atom        &  \ce{He} &  \ce{Li} &  \ce{Be}  &   \ce{B}  &   \ce{C}  &   \ce{N}  &  \ce{O}   &   \ce{F}  &   \ce{Ne}  \\
\hline
cc-pVDZ     & $-2.897\,5$ & $-7.477\,1$ & $-14.668\,1$ & $-24.645\,1$ & $-37.827\,5$ & $-54.556\,9$ & $-75.016\,3$ & $-99.658\,6$ & $-128.837\,5$ \\
cc-pVTZ     & $-2.903\,3$ & $-7.478\,1$ & $-14.668\,1$ & $-24.649\,9$ & $-37.839\,0$ & $-54.579\,5$ & $-75.052\,7$ & $-99.713\,9$ & $-128.909\,9$ \\
cc-pVQZ     & $-2.903\,7$ & $-7.478\,5$ & $-14.667\,6$ & $-24.652\,7$ & $-37.844\,3$ & $-54.588\,0$ & $-75.065\,8$ & $-99.732\,9$ & $-128.936\,0$ \\
Exact\fnm[1]& $-2.903\,7$ & $-7.478\,1$ & $-14.667\,4$ & $-24.653\,9$ & $-37.845\,0$ & $-54.589\,2$ & $-75.067\,3$ & $-99.733\,9$ & $-128.937\,6$
\end{tabular}
\end{ruledtabular}
\fnt[1]{Values extracted from Ref.~\onlinecite{ChaGwaDavParFro-PRA-93}.}
\end{table*}

\begin{table*}
\caption{
Total energies (in \si{\hartree}) obtained from TC-FCI calculations (without frozen-core approximation) in the cc-pVDZ, cc-pVTZ, and
cc-pVQZ basis sets for \ce{Li2}, \ce{Be2}, \ce{H2O}, \ce{NH3}, \ce{CH4}, and \ce{H2CO}.
}
\label{tab:ETC_total_molec}
\begin{ruledtabular}
\begin{tabular}{ccccccc}
Molecule    &  \ce{Li2} &  \ce{Be2} &  \ce{H2O}  &  \ce{NH3} &  \ce{CH4} &  \ce{H2CO} \\
\hline
cc-pVDZ     & $-14.977\,6$ & $-29.324\,5$ & $-76.377\,5$  & $-56.516\,2$ & $-40.503\,1$ & $-114.401\,2$ \\
cc-pVTZ     & $-14.988\,8$ & $-29.326\,8$ & $-76.423\,4$  & $-56.556\,6$ & $-40.521\,1$ & $-114.483\,5$ \\
cc-pVQZ     & $-14.991\,8$ & $-29.339\,5$ & $-76.436\,7$  & $-56.565\,5$ & $-40.520\,7$ & $-114.509\,7$
\end{tabular}
\end{ruledtabular}
\end{table*}

\begin{table}
	\caption{
	Total energies (in \si{\hartree}) computed at the FCI, TC-FCI, and FN-DMC levels of theory
	for the water molecule in the cc-pCVXZ family of basis sets.
	The estimated exact energy is \SI{-76.4389}{\hartree}. \cite{Klo-MP-01}
	}
	\label{tab:Etotal_water}
\begin{ruledtabular}
\begin{tabular}{cccc}
         & FCI\fnm[1]    & TC-FCI\fnm[2] & FN-DMC\fnm[3]   \\
\hline
cc-pCVDZ & $-76.282\,9$ & $-76.388\,4$ & $-76.415\,71(20)$ \\
cc-pCVTZ & $-76.390\,2$ & $-76.428\,6$ & $-76.431\,82(19)$ \\
cc-pCVQZ & $-76.421\,2$ & $-76.436\,5$ & $-76.436\,22(14)$ \\
cc-pCV5Z & $-76.431\,1$ &              & $-76.437\,44(18)$
\end{tabular}
\end{ruledtabular}
\fnt[1]{Values extracted from  Ref.~\onlinecite{AlmoraDiaz_2014}.}
\fnt[2]{This work.}
\fnt[3]{Values extracted from Ref.~\onlinecite{CafAplGinSce-JCP-16}.}
\end{table}

\begin{figure}[]
	\includegraphics[width=0.49\textwidth]{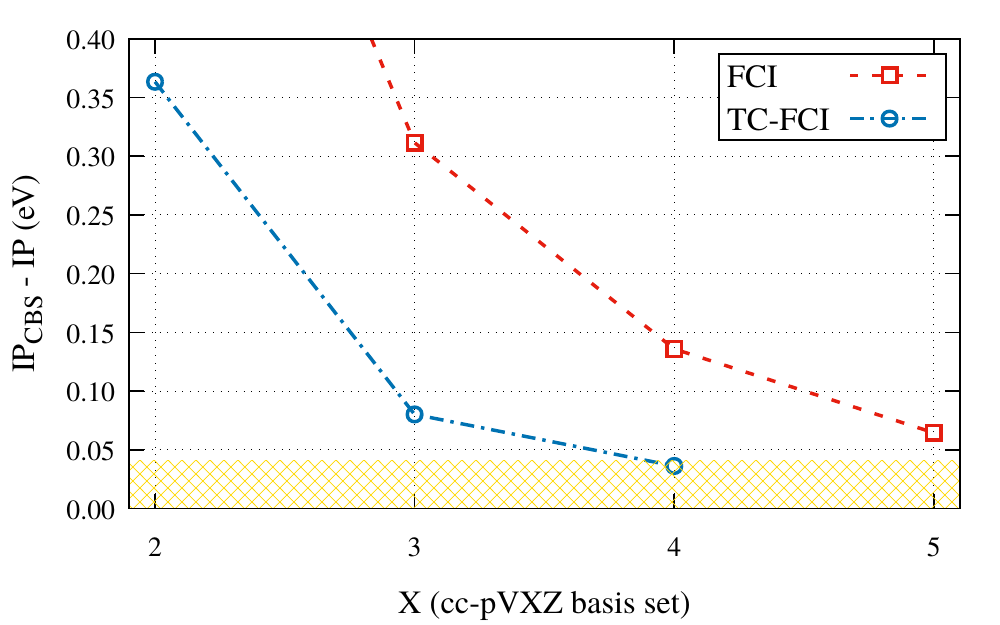}
	\caption{
	Error (in \si{\eV}) in the IP of \ce{H2O} computed at the FCI and TC-FCI
	levels with the cc-pVXZ family of basis sets with respect to the CBS estimate (\SI{12.80}{\eV}).
	The shaded yellow region corresponds to \SI{1.5}{\milli\hartree} accuracy.
	}
	\label{fig:IP_H2O}
\end{figure}

\subsection{Ionization Potentials}
\label{sec:IP}

In this section, we address the convergence of IPs (defined as the difference between the cation and neutral ground-state energies) as the one-electron basis set is enlarged.
Concerning the computation of IPs within the TC formalism, the optimal approach would be to optimize the Jastrow parameters in a state-specific way.
While not ideal, we have employed an alternative, cheaper strategy where the same Jastrow parameters are used for both the neutral and cationic systems.
(Here, we rely on the optimized parameters of the neutral system.)

We have computed the TC-FCI energies of the same set of atoms as in Sec.~\ref{sec:total_energies}, both with and without the frozen-core approximation, using the cc-pVXZ basis sets (where X = D, T, and Q).
The total energies are provided in \SupInf while the resulting IPs, expressed in \si{\eV}, are depicted in Table \ref{tab:IP_atoms}.
The absolute deviation caused by the frozen-core approximation is approximately \SI{1}{\milli\eV} on average.
This suggests that the frozen-core approximation can be effectively employed in TC-CI calculations to reduce the size of the Hilbert space without deteriorating the precision.
Furthermore, upon comparison with the exact IP estimates reported in Ref.~\onlinecite{ChaGwaDavParFro-PRA-93}, we observe mean absolute errors of \SI{0.22}{\eV}, \SI{0.02}{\eV}, and \SI{0.03}{\eV} for the cc-pVDZ, cc-pVTZ, and cc-pVQZ basis sets, respectively.
A slight increase of the error is observed between cc-pVTZ and cc-pVQZ, which we attribute to the imbalance in accuracy between the neutral and cation species. Indeed, in the cc-pVQZ basis, the total energy of the neutral systems is nearly exact while the energy of the cations is comparatively less precise.
Additionally, we report in Table \ref{tab:IP_atoms} the IPs computed at the FCI level for the same family of basis sets.
The mean absolute errors are \SI{0.37}{\eV}, \SI{0.14}{\eV}, \SI{0.06}{\eV}, and \SI{0.03}{\eV} for the cc-pVDZ, cc-pVTZ, cc-pVQZ, and cc-pV5Z basis sets, respectively.
In conclusion, while cc-pVTZ is sufficient to achieve chemical accuracy at the TC-FCI level, one must employ cc-pV5Z at the FCI level.

In Table \ref{tab:IP_molec}, we have conducted FCI and TC-FCI calculations, with or without the frozen-core approximation,
to estimate the IP of the following molecules: \ce{Li2}, \ce{Be2}, \ce{H2O}, \ce{NH3}, \ce{CH4}, and \ce{H2CO}.
Similar to the case of atoms, the mean absolute deviation induced by the frozen-core approximation is negligible (\SI{3}{\milli\eV}).
To produce CBS estimates, we have extrapolated the IPs obtained at the FCI level using an inverse cubic parametrization, across
the cc-pVXZ basis sets (where X = T, Q, and 5).
Considering the CBS estimates as references, the mean absolute errors are \SI{0.40}{\eV}, \SI{0.15}{\eV}, \SI{0.06}{\eV},
and \SI{0.03}{\eV} at the FCI level for the cc-pVDZ, cc-pVTZ, cc-pVQZ, and cc-pV5Z basis sets, respectively. These errors are
reduced to \SI{0.17}{\eV}, \SI{0.05}{\eV}, and \SI{0.02}{\eV} at the TC-FCI level for cc-pVDZ, cc-pVTZ, and cc-pVQZ, respectively.
Therefore, similar conclusions hold for the case of molecules.
To further illustrate this, we present, in Fig.~\ref{fig:IP_H2O}, the difference between the CBS estimate (\SI{12.80}{\eV}) and the IP values obtained at the FCI and TC-FCI levels across the cc-pVXZ basis sets in the case of the water molecule.
While chemical accuracy is nearly attained at the TC-FCI/cc-pVTZ level, the error is three times larger at the FCI/cc-pVTZ level.

\begin{table*}
\caption{
	IPs (in \si{\eV}) computed at the FCI and TC-FCI levels with or without frozen-core (FC) approximation for the neutral atoms from $Z=2$ to $Z=10$.
	For the sake of comparison, estimated exact IPs are also provided.}
\label{tab:IP_atoms}
\begin{ruledtabular}
\begin{tabular}{cccccccccc}
    &     \multicolumn{3}{c}{\ce{He}}    &     \multicolumn{3}{c}{\ce{Li}}   &     \multicolumn{3}{c}{\ce{Be}}   \\
        & FCI   & \multicolumn{2}{c}{TC-FCI} & FCI  & \multicolumn{2}{c}{TC-FCI} & FCI  & \multicolumn{2}{c}{TC-FCI} \\
        \cline{2-4} \cline{5-7} \cline{8-10}
cc-pVDZ & 24.33 & \multicolumn{2}{c}{24.60}  & 5.35 & \multicolumn{2}{c}{5.13}   & 9.29 & \multicolumn{2}{c}{9.29}   \\
cc-pVTZ & 24.53 & \multicolumn{2}{c}{24.61}  & 5.35 & \multicolumn{2}{c}{5.40}   & 9.29 & \multicolumn{2}{c}{9.32}   \\
cc-pVQZ & 24.56 & \multicolumn{2}{c}{24.59}  & 5.37 & \multicolumn{2}{c}{5.40}   & 9.30 & \multicolumn{2}{c}{9.33}   \\
cc-pV5Z & 24.58 &                 &          & 5.38 &                 &          & 9.31 &              &             \\
Exact\fnm[1] &   \multicolumn{3}{c}{24.59}   &    \multicolumn{3}{c}{5.39}       &      \multicolumn{3}{c}{9.32}     \\ \hline
    &     \multicolumn{3}{c}{\ce{B}}  &     \multicolumn{3}{c}{\ce{C}}  &     \multicolumn{3}{c}{\ce{N}}  \\
        & FCI (FC) & TC-FCI (FC) & TC-FCI & FCI (FC) & TC-FCI (FC) & TC-FCI & FCI (FC) & TC-FCI (FC) & TC-FCI \\ \cline{2-4} \cline{5-7} \cline{8-10}
cc-pVDZ &  8.07    &  8.15       &  8.15  &  10.98   &  11.08      &  11.08 &  14.19   &  14.33      &  14.33 \\
cc-pVTZ &  8.22    &  8.31       &  8.31  &  11.17   &  11.27      &  11.28 &  14.43   &  14.56      &  14.56 \\
cc-pVQZ &  8.24    &  8.30       &  8.30  &  11.21   &  11.27      &  11.27 &  14.49   &  14.56      &  14.56 \\
cc-pV5Z &  8.25    &             &        &  11.22   &             &        &  14.51   &             &        \\
Exact\fnm[1] &   \multicolumn{3}{c}{8.30} &    \multicolumn{3}{c}{11.26}    &     \multicolumn{3}{c}{14.53}   \\ \hline
    &  \multicolumn{3}{c}{\ce{O}}     &     \multicolumn{3}{c}{\ce{F}}  &     \multicolumn{3}{c}{\ce{Ne}} \\
        & FCI (FC) & TC-FCI (FC) & TC-FCI & FCI (FC) & TC-FCI (FC) & TC-FCI & FCI (FC) & TC-FCI (FC) & TC-FCI \\ \cline{2-4} \cline{5-7} \cline{8-10}
cc-pVDZ &  12.85   &  13.13      & 13.13  &  16.71   &  17.09      & 17.08  &  20.89   &  21.22      &  21.22 \\
cc-pVTZ &  13.32   &  13.58      & 13.58  &  17.13   &  17.45      & 17.45  &  21.30   &  21.62      &  21.62 \\
cc-pVQZ &  13.48   &  13.63      & 13.62  &  17.30   &  17.49      & 17.48  &  21.47   &  21.67      &  21.67 \\
cc-pV5Z &  13.55   &             &        &  17.37   &             &        &  21.54   &             &        \\
Exact\fnm[1] & \multicolumn{3}{c}{13.62}  &      \multicolumn{3}{c}{17.42}  &     \multicolumn{3}{c}{21.56}
\end{tabular}
\end{ruledtabular}
\fnt[1]{Values extracted from Ref.~\onlinecite{ChaGwaDavParFro-PRA-93}.}
\end{table*}

\begin{table*}
\caption{
	IPs (in \si{\eV}) computed at the FCI and TC-FCI levels with or without frozen-core (FC) approximation for \ce{Li2}, \ce{Be2}, \ce{H2O}, \ce{NH3}, \ce{CH4}, and \ce{H2CO}.
	The CBS estimates are obtained by extrapolation based on the FCI data.
}
\label{tab:IP_molec}
\begin{ruledtabular}
\begin{tabular}{cccccccccc}
        &     \multicolumn{3}{c}{\ce{Li2}}   &     \multicolumn{3}{c}{\ce{Be2}}   &  \multicolumn{3}{c}{\ce{H2O}}   \\
        &  FCI  & \multicolumn{2}{c}{TC-FCI} &  FCI  & \multicolumn{2}{c}{TC-FCI} & FCI (FC) & TC-FCI (FC) & TC-FCI \\ \cline{2-4} \cline{5-7} \cline{8-10}
cc-pVDZ &  5.19 & \multicolumn{2}{c}{5.18}   &  7.44 & \multicolumn{2}{c}{7.47}   &  11.96   &    12.44    &  12.43 \\
cc-pVTZ &  5.22 & \multicolumn{2}{c}{5.26}   &  7.45 & \multicolumn{2}{c}{7.50}   &  12.49   &    12.72    &  12.71 \\
cc-pVQZ &  5.24 & \multicolumn{2}{c}{5.27}   &  7.47 & \multicolumn{2}{c}{7.50}   &  12.66   &    12.76    &  12.76 \\
cc-pV5Z &  5.25 &              &             &  7.48 &              &             &  12.73   &             &        \\
CBS     &  5.25 &              &             &  7.49 &              &             &  12.80   &             &        \\ \hline
        &  \multicolumn{3}{c}{\ce{NH3}}   &  \multicolumn{3}{c}{\ce{CH4}}   &   \multicolumn{3}{c}{\ce{H2CO}}   \\
        & FCI (FC) & TC-FCI (FC) & TC-FCI & FCI (FC) & TC-FCI (FC) & TC-FCI &  FCI (FC)  & TC-FCI (FC) & TC-FCI \\ \cline{2-4} \cline{5-7} \cline{8-10}
cc-pVDZ &  10.32   &    10.68    & 10.68  &  14.20   &    14.42    & 14.40  &   10.44    &   10.76     &  10.76 \\
cc-pVTZ &  10.76   &    10.92    & 10.91  &  14.36   &    14.47    & 14.47  &   10.80    &   10.95     &  10.95 \\
cc-pVQZ &  10.89   &    10.96    & 10.96  &  14.40   &    14.44    & 14.44  &   10.92    &   10.98     &  10.98 \\
cc-pV5Z &  10.95   &             &        &  14.41   &             &        &   10.95    &             &        \\
CBS     &  11.00   &             &        &  14.43   &             &        &   11.00    &             &
\end{tabular}
\end{ruledtabular}
\end{table*}

\section{Conclusion}
\label{sec:conclusion}

In the present study, we investigated how incorporating a correlation factor in the Hamiltonian within the TC framework markedly accelerates the convergence of SCI methods. 
To illustrate these results, we systematically studied the total energies and IPs in increasing large Dunning basis sets 
using optimized Jastrow factors available from the literature in the case of neutral atoms with $2 \le Z \le 10$ together with the \ce{Li2}, \ce{Be2}, \ce{H2O}, \ce{NH3}, \ce{CH4}, and \ce{H2CO} molecules.
This acceleration has been demonstrated quantitatively through both the reduction in the size of the one-electron basis 
functions and the decrease in the number of important determinants in the Hilbert space.

Although the TC Hamiltonian exhibits certain difficulties related to three-electron terms and non-Hermiticity, the 
working equations derived here for TC-SCI demonstrate a scaling similar to that of SCI methods.
Nonetheless, as our numerical analysis reveals, TC-SCI effectively increases the sparsity of the Slater determinant 
space when a flexible enough Jastrow factor is employed.
This enhancement enables faster and more stable convergence towards the FCI limit within a given basis set.

%
By comparing with the best estimates (when available for total energies) or CBS values reported here for IPs, we 
have shown that achieving near-exact results, \textit{i.e.} within chemical accuracy, is possible in TC-SCI 
using basis sets significantly smaller than those required in standard SCI calculations.

Expanding the applicability of TC-SCI methods to arbitrary systems necessitates the ability to optimize 
Jastrow factors at a reasonable cost. 
While this challenge has been considered in prior studies, it is widely acknowledged to be both difficult and expensive. 
In our upcoming efforts, we will explore how to systematically optimize Jastrow factors at a reasonable cost.

\acknowledgments{
This work was performed using HPC resources from GENCI-TGCC
(gen1738,gen12363) and from CALMIP (Toulouse) under allocation
2024-18005, and was also supported by the European Centre of
Excellence in Exascale Computing TREX --- Targeting Real Chemical
Accuracy at the Exascale. This project has received funding from the
European Union's Horizon 2020 --- Research and Innovation program ---
under grant agreement no.~952165.
A.A., A.S., and P.F.L.~also acknowledge funding from the European Research Council (ERC) under the European Union’s Horizon 2020 research and innovation programme (Grant agreement No.~863481).
}

\section*{Data availability statement}
The data that supports the findings of this study are available within the article and its supplementary material.

\section*{References}

\bibliography{main}

\end{document}